\documentclass[useAMS,usenatbib]{mn2e}
\usepackage{graphicx}
\usepackage{graphicx,xspace}
\usepackage{epsf}
\usepackage{amsmath}
\usepackage{color}
\usepackage{epsfig,epstopdf}

\newcommand\T{\rule{0pt}{2.6ex}}   
\def\gsim { \lower .75ex \hbox{$\sim$} \llap{\raise .27ex \hbox{$>$}} }
\def\lsim { \lower .75ex \hbox{$\sim$} \llap{\raise .27ex \hbox{$<$}} }

\newcommand{\FoF}{FoF\xspace}

\newcommand{\subfind}{\textsc{subfind}\xspace}

\newcommand{\aap}{A\&A}

\newcommand{\apjl}{ApJ}
\newcommand{\apjs}{ApJS}

\renewcommand\tabcolsep{0.2cm}

\title[satellite orbital parameters]  
{Orbital parameters of infalling 
satellite haloes in the hierarchical $\Lambda$CDM model}
\author[Jiang et al.] {\parbox{18cm}{
Lilian Jiang, Shaun Cole, Till Sawala, Carlos S.~Frenk
}\vspace{0.3cm}\\
{Institute for Computational Cosmology, Dept. of Physics, Univ. of
  Durham, South Road, Durham  DH1 3LE, UK}
  }

\begin{document}
\pagerange{\pageref{firstpage}--\pageref{lastpage}} \pubyear{2014}
\maketitle
\label{firstpage}


\begin{abstract} {We present distributions of orbital parameters of
    infalling satellites of $\Lambda$CDM haloes in the mass range
    $10^{12}-10^{14}$M$_\odot$, which represent the initial conditions
    for the subsequent evolution of substructures within the host
    halo. We use merger trees constructed in a high resolution
    cosmological N-body simulation to trace satellite haloes, and
    identify the time of infall. We find signficant trends in the
    distribution of orbital parameters with both the host halo mass
    and the ratio of satellite-to-host halo masses. For all host halo
    masses, satellites whose infall mass is a larger fraction of the
    host halo mass have more eccentric, radially biased orbits. At
    fixed satellite-to-host halo mass ratio, high mass haloes are
    biased towards accreting satellites on slightly more radial
    orbits. To charactise the orbital distributions fully requires
    fitting the correlated bivariate distribution of two chosen
    orbital parameters (e.g.  radial and tangential velocity or energy
    and angular momentum). We provide simple fits to one choice of the
    bivariate distributions, which when transformed faithfully,
    captures the behaviour of any of the projected one-dimensional
    distributions.}
\end{abstract}


\section{INTRODUCTION}

In the current cosmological structure formation model, dark matter
haloes grow by the merging of smaller systems \citep{white78,davis85},
leading to hierarchical halo growth. Substructures that are accreted
onto a host halo can survive for significant periods of time within
the host halo \citep{ch43,klypin99,moore99,bin08,bk08,jiang08}.  These
substructures can host satellite galaxies, such as those found in the
Local Group, and galaxy clusters. Thus, it is important to study the
distribution of the initial orbital parameters of subhaloes at the
time of infall as they represent the initial conditions which
determine the later evolution of the substructures in their host
haloes.

Semi-analytic models of galaxy formation rely on prescriptions for
dynamical friction survival times and tidal stripping, (see
\citealt{baugh06} for a review). Assuming the halo potential to be
spherically symmetric, a satellite orbit can be defined by the plane
of the orbit and two further parameters related to the energy and
angular momentum such as circularity and pericentre.  Previous authors
have studied the distributions of such orbital parameters for
substructures in numerical simulations \citep{tormen97, vit02,
  benson05, wang05, zentner05, kandb06, wetzel11}. \cite{tormen97}
investigated the infall of satellites into the haloes of galaxy
cluster mass, and reported that more massive satellites move along
slightly more eccentric orbits, with lower specific angular momentum
and smaller pericentres.  \cite{benson05} presented evidence for a
satellite mass dependence of the distribution of orbital parameters,
but was unable to characterise these trends accurately due to the
limited statistics. Apparently in slight contradiction,
\cite{wetzel11} reports that the orbital parameters do not
significantly depend on the satellite halo mass but depend more on the
host halo mass.  These studies were hampered by limited dynamic range
and sample size. The high resolution and large volume of the
simulation we analyse allow us to quantify trends in both satellite
and host halo mass.

The two parameters characterising a satellite orbit are, in general,
correlated.  \cite{wetzel11} provides fits to circularity and
pericentre, but he stopped short of examining correlations between
these parameters which are important if one wants to select
representative orbits from the distribution.  \cite{kandb06} found a
tight correlation between pericentre and
circularity. \cite{tormen97,gill04,benson05} also find correlations
between orbital parameters.

In this paper, we investigate the correlations between different
possible pairs of parameters.  We show that that to a good
approximation total infall velocity and the fraction of this velocity
which is in the radial direction are uncorrelated. We present fits to
these and show that when transformed these fits provide accurate
descriptions of the distributions of other choices of orbital
parameters.

Most previous work has focused on orbits only at redshift $z=0$, or on
the satellites that are still identified at $z=0$. In our work we
focus on host haloes that exist at $z=0$, but we analyse the orbits of
all satellites that fall into the host halo after its formation
(defined as when its main progenitor had half the final halo mass),
regardless of whether the satellite is still identifiable at $z=0$.
 
Our paper is structured as follows. In Section~2, we briefly outline
the methods including a detailed description of the N-body simulation, 
the identification of halo mergers and the measurement of orbital
parameters. In Section~3, we present detailed analysis of the orbital 
parameters. We conclude in Section~4. 

\section{Methods}
\label{sec:2}
\subsection{Simulation}
\label{sec:Dove}

Our analysis is based on the \textsc{DOVE} simulation, a $\Lambda$CDM 
cosmological dark matter only simulation of a periodic volume
  with side length $100$~Mpc, with cosmological parameters adapted
  from the \textsc{wmap7} analysis of \cite{komatsu2011}. The Hubble
parameter, density parameter, cosmological constant, scalar spectral
index and linear rms mass fluctuation in 8~$h^{-1}$Mpc radius spheres
were $H_0= 70.4$~km$\,$s$^{-1}$, $\Omega_{\rm m}=0.272$,
$\Omega_\Lambda=0.728$, $n_{\rm s}=0.97$ and $\sigma_8=0.81$,
respectively. The dark matter is represented by $N_{\rm p}=1620^3$
particles of mass $m_{\rm p} = 8.8\times10^6$~M$_\odot$.  Initial
conditions were set up using second order Lagrangian perturbation
theory \citep{jenkins2010}, with phases set using the multiscale
Gaussian white noise field {\it Panphasia} \citep{jenkins2013}.  These
phases were chosen to be the same as in the {\sc eagle} simulation
\citep{schaye14} and are fully specified by the {\it Panphasia}
descriptor [Panph1,L16,(31250,23438,39063),S12,CH1050187043,
EAGLE\_L0100\_VOL1].  The initial conditions were evolved to $z=0$
using the {\sc gadget3} N-body code, which is an enhanced version of
the code described in \cite{springel2005}.

The particle positions and velocities were output at 160
  snapshots, equally spaced in $\log(a)$ from $z=20$. At each output,
  haloes were identified using a Friends-of-Friends algorithm
  \citep[FoF;][]{davis85}, and the {\sc subfind} algorithm
  \citep{springel01} was used to identify self-bound substructures
  (``subhaloes'') within them. We define our \FoF haloes by the
  conventional linking length parameter of $b=0.2$ (the linking length
  is defined as $b$ times the mean interparticle
  separation). Typically the main \subfind subhalo contains most of
  the mass of the original \FoF halo, only unbound particles and those
  bound to secondary subhaloes are excluded. We keep all haloes and
  subhaloes with more than 20 particles, corresponding to
  $2\times10^8$~M$_\odot$.

\subsection{Orbital Parameters}
\label{sub:method}
We define the virial mass, $M_{\rm vir}$, and associated virial
radius, $r_{\rm vir}$, of a dark matter halo using a simple spherical
overdensity criterion centred on the potential minimum:
\begin{equation}
M_{\rm vir} =\frac{4}{3}\pi \Delta \, \rho_{\rm crit} \,
r_{\rm vir}^3
\label{eq:rvir}
\end{equation}
where $\rho_{\rm crit}$ is the cosmological critical density and
$\Delta$ is the specified overdensity. We adopt $\Delta=200$ and
include all the particles inside this spherical volume, not only the
particles grouped by the adopted halo finder, to define the enclosed
mass, $M_{200}$, and associated radius $r_{200}$.  This choice of
$\Delta=200$ is largely a matter of convention, but has been shown
roughly to correspond to the boundary at which the haloes are in
approximate dynamical equilibrium \cite[e.g.][]{cole96}.  We express
velocities in units of the virial velocity, $V_{200}$, of the host
halo.

For a spherical potential, the orbit of a satellite can be fully
specified by the orientation of the orbit and two non-trivial
parameters related to its energy, $E$, and the modulus of its angular
momentum, $J$. There are various choices for these two parameters.
The choice made by \cite{benson05} and others of the radial, $V_r$,
and tangential, $V_\theta$, velocities at infall benefits from being
directly measurable quantities and being simple.  In contrast,
\cite{tormen97} adopted the \emph{circularity}, defined as the total
angular momentum in units of the angular momentum for a circular orbit
of the same energy, $J/J_{\rm circ}(E)$, and the infall radius in
units of the radius of a circular orbit of the same energy, $r/r_{\rm
  circ}(E)$. These have the advantage of depending only on the
conserved quantities $E$ and $J$ (Note, the $r$ here is the radius at
infall and so equals $r_{200}$ in our study.), but require adopting a
model of the halo potential. The particular form of these two
parameters is motivated by theoretical modelling including that of
satellite orbital decay due to dynamical friction
\citep{lacey93,jiang08}. To define these two parameters, we adopt a
singular isothermal sphere (SIS) \citep{cole96} as a simple model for
the density profile of dark matter haloes. 
This choice is consistent
with assumptions in \cite{lacey93} and so provides orbital
parameters that can be directly substituted into their merger
time formula. 
In Section~\ref{sec:derived} we also provide formulae for computing the 
corresponding orbital parameters if one instead adopts a more realistic
NFW potential \citep{1996ApJ...462..563N}.

Here we derive the transformations between these two parametrisations.
Defining the zero point of the gravitational
potential to be at $r_{200}$, where the circular velocity, $V_{200}$,
is given by $V_{200} = \sqrt{GM_{200}/r_{200}}$, we can express the
gravitational potential as
\begin{equation}
\phi(r)  = V^2_{200} \ln(r/r_{200}) .
\end{equation}
Thus, for a satellite crossing $r_{200}$ with radial and tangential
velocities, $V_r$ and $V_\theta$, the total energy per unit mass is
\begin{equation}
E  = \frac{1}{2} \left( V^2_r + V_\theta^2 \right).
\end{equation}
As the circular velocity is constant for a SIS, the radius, of a
circular orbit of the same energy is given by
\begin{equation}
\frac{1}{2} \left( V^2_r + V_\theta^2 \right) 
= \frac{1}{2} V_{200}^2+ V^2_{200} \ln(r_{\rm circ}/r_{200}),
\end{equation}
implying
\begin{equation}
\frac{r_{\rm circ}(E)}{r_{200}} =  \exp\left(\frac{V^2_r + V_\theta^2 - V_{200}^2 }
{2 V_{200}^2}\right). 
\label{equation:rcr200}
\end{equation}
As the corresponding angular momentum of a circular orbit is $J_{\rm
  circ}(E)=M_{\rm s} V_{200} r_{\rm circ}(E)$, we have
\begin{equation}
\frac{J}{J_{\rm circ}(E)} =\frac{V_\theta}{V_{200}}
\exp\left( -\frac{V^2_r + V_\theta^2 - V_{200}^2 }
{2 V_{200}^2}\right) . 
\end{equation}

Another useful quantity to define is the composite parameter
\begin{equation}
\Theta= \left(\frac{J}{J_{\rm circ}(E)}\right)^{0.78}\left(\frac{r_{200}}{r_{\rm circ}(E)}\right)^{2}. 
\label{eq:Theta}
\end{equation}
Its utility is that  \cite{lacey93} showed that the 
orbital decay time of a satellite of mass $M_{\rm s}$
due to dynamical friction within a host halo of mass
$M_{\rm h}$ is given by
\begin{equation}
\tau_{\rm mrg} =  \Theta\, \tau_{\rm dyn}
\frac{0.3722}{\ln(\Lambda_{\rm coulomb} )}\frac{M_{\rm
  h}}{M_{\rm s}} ,
\label{eq:taumrg}
\end{equation}
where $\tau_{\rm dyn}$ is the dynamical time of the host halo and
$\ln(\Lambda_{\rm coulomb} )$ is taken to be $\ln({M_{\rm h}}/{M_{\rm
    s}})$. This formula assumes that the satellite can be treated as a
point mass orbiting in a host halo with a SIS density profile and is
valid when $\tau_{\rm mrg} \gg \tau_{\rm dyn}$.  In this model it is
only necessary to know the one-dimensional distribution of $\Theta$
values rather than the bivariate distribution of, say, $V_r$, and
$V_\theta$ to determine the distribution of orbital decay times.

\begin{figure}
\resizebox{8.5cm}{!}{\includegraphics{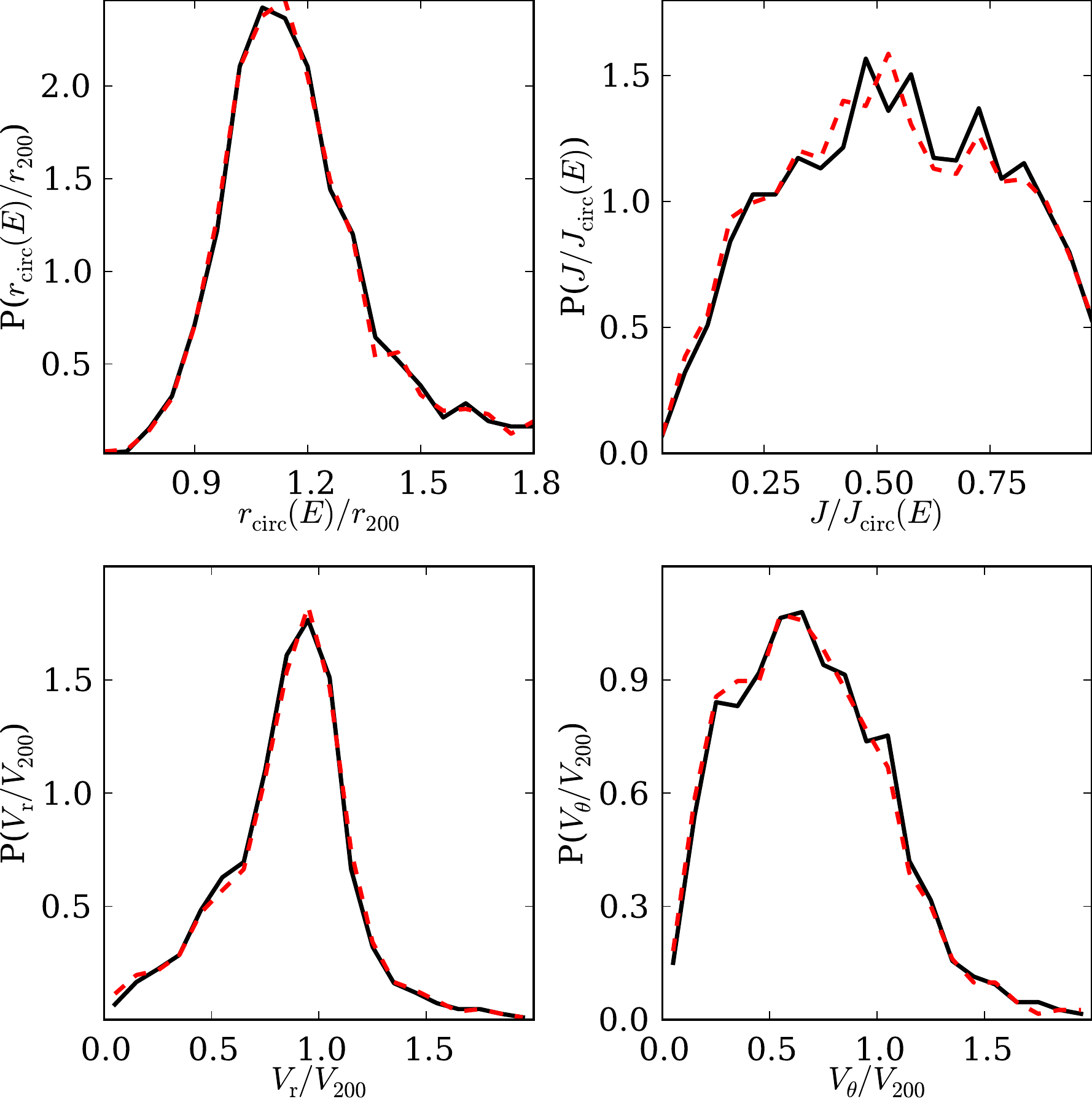}}%
\caption{Tests of the interpolation scheme on the distributions of the 
orbital parameters $r_{\rm
circ}(E)/r_{\rm 200}$, $J/J_{\rm circ}(E)$, $V_r/V_{\rm 200}$ and $V_{\rm
\theta}/V_{\rm 200}$. The panel shows the differential distribution of
orbital parameters in the mass ratio bin: $M_{\rm s}/
M_{\rm h}>0.05$ for all the host haloes in our sample.  Solid
lines show the results using linear interpolation of energy and
angular momentum, dotted lines show results using linear interpolation
of velocity and position.}
\label{fig:method}
\end{figure}

\subsection{Identifying halo mergers}
\label{sec:Trees}
We follow the evolution, infall and merging of haloes and
  subhaloes using merger trees. Our starting point is the catalogue
of \FoF haloes and their constituent subhaloes at redshift zero. We
build subhalo merger trees linking each subhalo to its progenitors and
descendants using the algorithm described in Appendix~A2 of
\cite{jiang14}. Next, we identify both the progenitors of the \FoF
haloes and the subhaloes which fall into them.  For each \FoF halo, we
trace its progenitor in the previous snapshot by identifying the main
progenitor of its main subhalo.  We then define the virial radius of
this progenitor halo such that a sphere of this radius centred on the
particle at the potential minimum of the main subhalo encloses 200
times the critical density as defined in Eqn.~\ref{eq:rvir}. We trace
the main progenitor of each redshift zero \FoF halo back in this way
until the last snapshot at which its mass is greater than half the
final halo mass. We choose not to consider mergers before the formation 
time of the main halo as we bin our results by the halo mass at $z=0$ and wish
this to relect (within a factor of two) the mass of the main halo when the
merger takes place. To identify subhaloes that merge onto this main
halo progenitor we not only trace the progenitors of subhaloes that
are in the halo at redshift zero, but also those that were inside
progenitors of the main halo at some point but which have since been
disrupted, merged or escaped. Hence, we trace every individual
subhalo from its formation redshift to the redshift when it first
crosses the virial radius of the host halo.

In order to find the precise crossing time, we save the orbital
information from the snapshots just before and after a satellite
subhalo crosses the virial radius. Then, we interpolate both the
satellite position (relative to the halo centre) and the halo virial
radius linearly to find the time when the subhalo first crosses the
virial radius.  To investigate the accuracy of the interpolation
scheme we considered two methods of interpolating the satellite
orbital parameters to this crossing time:
\begin{enumerate}
\item
  {We interpolate the energy (using the singular isothermal
  sphere approximation of the halo potential described in 
  Section~\ref{sub:method}) and angular momentum linearly in redshift
  to the crossing time. We then compute other orbital parameters
  such as the radial and tangential velocities from this interpolated
  energy and angular momentum.}
 \item 
 {Alternatively, we interpolate each component of the satellite's
  velocity  linearly in redshift  to the crossing time and
  then compute the required orbital parameters from the interpolated
  velocity and position.}
 \end{enumerate}
 Provided our simulation snapshots are sufficiently closely spaced, we
 would expect these two methods to give very similar results.  This is
 indeed what we find as demonstrated in Fig.~\ref{fig:method} which
 compares the distribution of the various orbital parameters for
 satellites satisfying $M_{\rm s}/M_{\rm h}>0.05$ at the time of
 infall in our full sample of haloes.  Throughout the rest of this
 paper, we show results just from the method that linearly
 interpolates the energy and angular momentum.  We would expect this
 to be the more accurate method as these two quantities are almost
 conserved and so only vary slowly with the interpolation parameter.

 Accurately defining the orbital parameters at the crossing time is an
 important issue that has been considered in earlier work. The
 approach adopted by \cite{benson05} and \cite{vit02} was to search
 for pairs of haloes within some separation $r_{\rm max}$ which are
 about to merge and then predict their crossing time by modelling them
 as two isolated point masses. A similar approach was taken by
 \cite{tormen97}, \cite{kandb06} and \cite{wetzel11}.  When using such
 schemes one must apply a weighting to correct for the
 under-representation of satellites with large infall velocities, some
 of which will be at separations greater than $r_{\rm max}$ at the
 earlier snapshot. In our work, due to the higher time resolution of
 our simulation outputs, we do not have to limit the separation
 between satellite and host halo at the snapshot prior to infall and
 instead form a complete census of all the infalling satellites.

\begin{figure}
\resizebox{8.5cm}{!}{\includegraphics{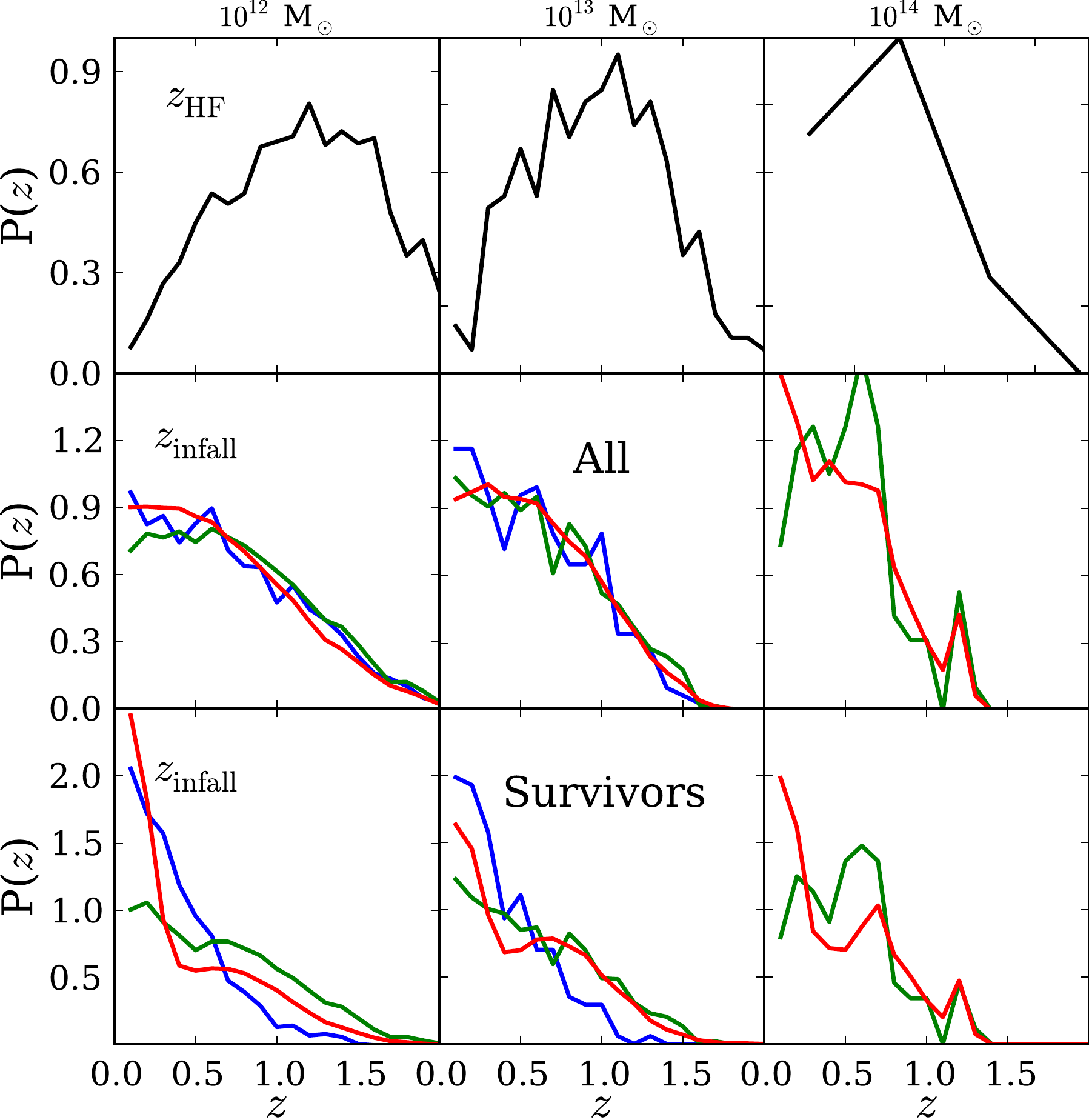}}%
\caption{The distributions of halo formation redshifts and
the redshifts at which satellites fall into these halos.
Each column is for a fixed final halo mass as labelled
at the top of the figure. The top row is the distribution
of halo formation redshifts. The middle row is the distribution
of satellite infall redshifts for all infalling satellites,
while bottom row is for the subset of these satellites which
survive as subhaloes at $z=0$. In the bottom two rows
the line colour indicates the satellite-to-host mass ratio.
The red lines are for $0.0001<M_{\rm s}/M_{\rm h}<0.005$,
green for $0.005<M_{\rm s}/M_{\rm h}<0.05$
and blue for $M_{\rm s}/M_{\rm h}>0.05$.
}
\label{fig:figz}
\end{figure}

\begin{figure}
\resizebox{8.5cm}{!}{\includegraphics{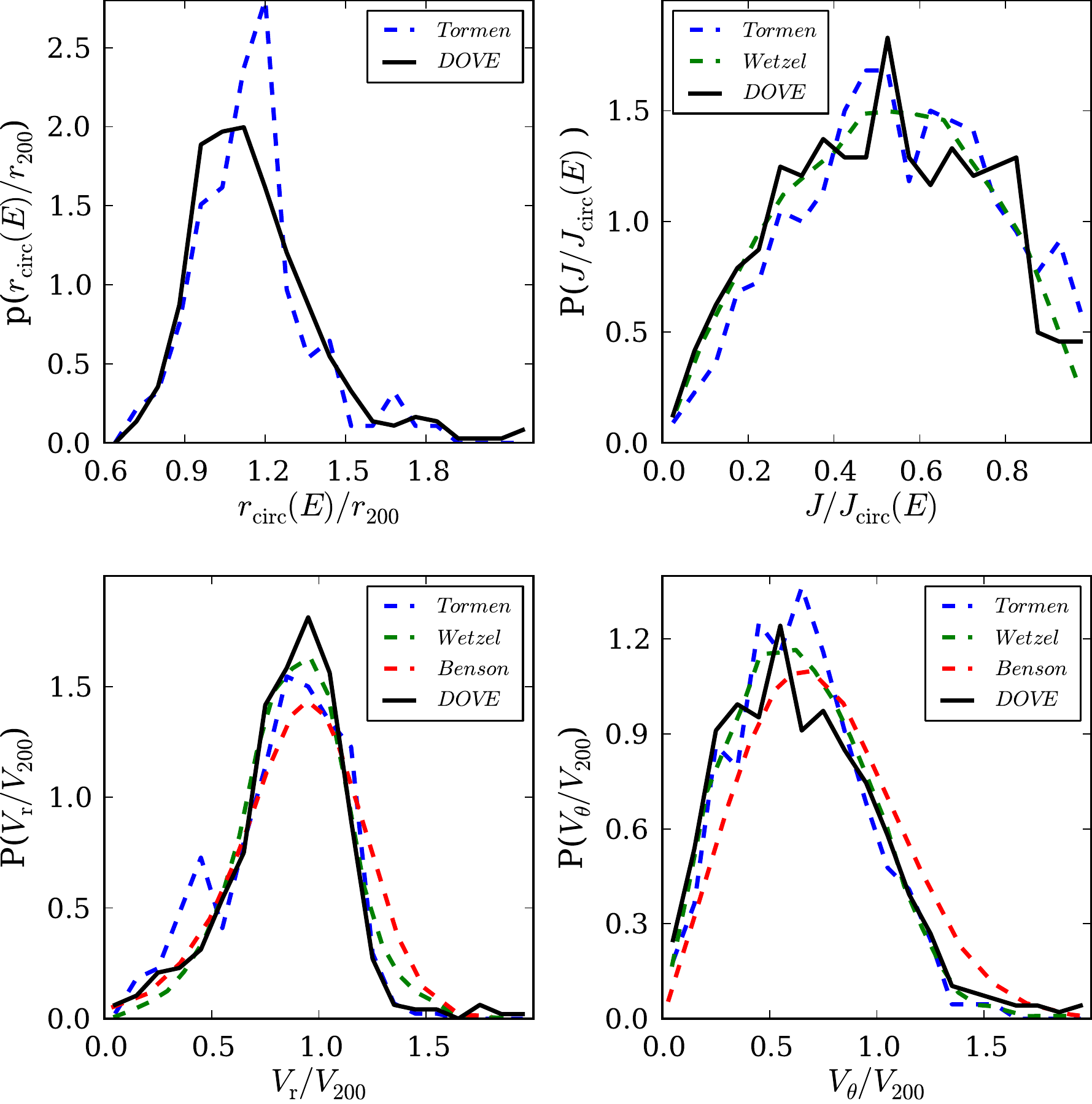}}
\caption{Comparison to published distributions of the orbital
  parameters $r_{\rm circ}(E)/r_{\rm 200}$, $J/J_{\rm circ}(E)$,
  $V_{\rm r}/V_{\rm 200}$, and $V_{\rm \theta}/V_{\rm 200}$.  In all
  the panels the black solid line shows the distribution of the
  satellite orbital parameters for infalling satellites in our
  analysed host haloes (covering the mass range $5\times10^{11}$ to
  $2.5\times10^{14}$~M$_{\rm \odot}$) with satellite-to-host halo mass
  ratios spanning $0.05$ to $0.5$. This range is typical of that
  probed by the samples to which we are comparing.  Blue, green and
  red dashed lines show the results from the work of Tormen (1997),
  Wetzel (2011) and Benson (2005) respectively.  }
\label{fig:fig_compare}
\end{figure}

\subsection{Formation and infall redshifts}

As we want our measured orbital parameter distributions to be
directly applicable to semi-analytic galaxy formation models 
we trace all the infalling subhaloes back to the formation time
of the main halo, where its formation time is defined
as when its main progenitor has half the final, $z=0$, halo mass.
We bin our halo samples by their mass at redshift $z=0$
and so by not tracing haloes back further in time we avoid 
significant ambiguity in the mass of the main halo at the 
time satellites are accreted, i.e. at all infall events 
the main halo is always within a factor of two the final halo mass.
The probability distribution function
of halo formation redshifts, $z_{\rm HF}$, (normalized such
that the integral over the distribution is unity)
are shown in the top row of Fig.~\ref{fig:figz} for each of our final
halo mass bins. 
As expected we see that lower mass haloes form earlier.
The median formation redshift of our $10^{12}$, $10^{13}$ 
and~$10^{14}$~M$_\odot$ haloes are 1.14, 0.92, and 0.66
respectively.

The middle row of Fig.~\ref{fig:figz} shows the
distribution of infall redshifts, $z_{\rm infall}$, split both
by final halo mass and by the ratio of satellite-to-host mass 
at infall. These distributions rise steadily towards redshift $z=0$
(though the distributions for the highest mass bin are noisy
  because of the limited size of that sample)
from the upper redshift set by when the first haloes in the sample
form. The most interesting aspect is that infall redshift distribution
at fixed halo mass is essentially independent of satellite-to-host 
mass ratio. This is equivalent to the mass distribution of the
infalling satellites, measured
in units of the host halo mass, being independent of redshift.
Given that the distribution of host halo masses is constrained
not to vary greatly with redshift (only haloes with mass greater than
half the final mass are retained in the sample) then this behaviour
is expected in simple excursion set models of hierarchical growth
\citep{lacey93}. 

 \begin{table*}
\caption{ The median values of infall redshifts
     for both all and surviving subhaloes and the orbital
    parameter, $\log_{10} \Theta$ for bins of final halo mass, $M_{\rm h}$, and
    the satellite-to-host mass ratio at infall, $M_{\rm s}/ M_{\rm
    h}$. The errors were estimated by bootstrap resampling of the halo
    sample.}
\label{table}
\begin{center}
\begin{tabular}{|c |c |c |c |l |}
\hline
  $M_{\rm h }$ &$M_{\rm s}/M_{\rm h}$ & $\log_{10} \Theta$& $z_{\rm infall}$~survivors&  $z_{\rm infall}$~all
   \hskip 1cm \T \\ \hline
10$^{12}$ M$_{\rm \odot }$
&$0.0001-0.005$&$\hphantom{-}0.076\pm{0.002}$&$0.217\pm{0.006}$&
$0.491\pm{0.007}$ \T \\
10$^{13}$ M$_{\rm \odot}$
&$0.0001-0.005$&$\hphantom{-}0.051\pm{0.004}$&$0.428\pm{0.025}$&
$0.516\pm{0.017}$\\
10$^{14}$ M$_{\rm \odot}$ &$0.0001-0.005$&$\hphantom{-}0.050\pm{0.017}$
&$0.380\pm{0.064}$&$0.409\pm{0.053}$ \\ \hline
10$^{12}$ M$_{\rm \odot}$&$0.005-0.05$ &$\hphantom{-}0.027\pm{0.003}$ &
$0.462\pm{0.009}$&$0.521\pm{0.010}$ \T \\
10$^{13}$ M$_{\rm \odot}$&$0.005-0.05$& ${-}0.009\pm{0.012}$&
$0.480\pm{0.024}$&$0.522\pm{0.022}$\\
10$^{14}$ M$_{\rm \odot}$&$0.005-0.05$& ${-}0.015\pm{0.047}$ &
$0.483\pm{0.086}$&$0.483\pm{0.062}$\\ \hline
10$^{12}$ M$_{\rm
\odot}$&$0.05-0.5$&${-}0.082\pm{0.007}$&$0.290\pm{0.012}$&$0.511\pm{0.019}$
\T \\
10$^{13}$ M$_{\rm \odot}$&$0.05-0.5$&
${-}0.130\pm{0.021}$&$0.268\pm{0.031}$&$0.484\pm{0.040}$

   \\\hline
\end{tabular}
\end{center}
\end{table*}

The bottom row of Fig.~\ref{fig:figz} also shows distributions
of infall redshifts, but now just for the satellites that survive
and are identifiable at redshift $z=0$. The median redshifts of these
distributions are compared to those of corresponding complete samples
in Table~\ref{table}. Comparing these values and the distributions
shown in the bottom and middle rows of Fig.~\ref{fig:figz} one clearly 
sees that the typical infall redshift of surviving satellites
is significantly lower than that of the complete sample.

This is, at least in part, a resolution effect as we are unable to
identify satellites with fewer than 20~particles. Thus the shift
to lower infall redshifts is greatest for the lowest mass satellites
which are the ones with the smaller satellite-to-host mass  ratio
in the lower halo mass bins.

\section{Orbital Parameter Distributions}
  
\subsection{Comparison to previous work}
Fig.~\ref{fig:fig_compare} compares our orbital parameter
distributions with those from \cite{tormen97}, \cite{benson05} and
\cite{wetzel11}.  In all panels, the black solid lines show the
distributions for satellites with mass ratios in the range
$0.05<M_{\rm s}/M_{\rm h}<0.5$ averaged over all our analysed haloes
which span the mass range $5\times10^{11}< M_{\rm
  h}<2.5\times10^{14}$M$_{\rm \odot}$.  In general our results are in
good agreement with these published datasets and those of
\cite{wang05, zentner05, kandb06}, despite variations between these
studies in the definition of crossing time and the choice of
cosmology.

\begin{figure*}
\begin{center}
\begin{tabular}{cc}
\resizebox{15.0cm}{!}{\includegraphics{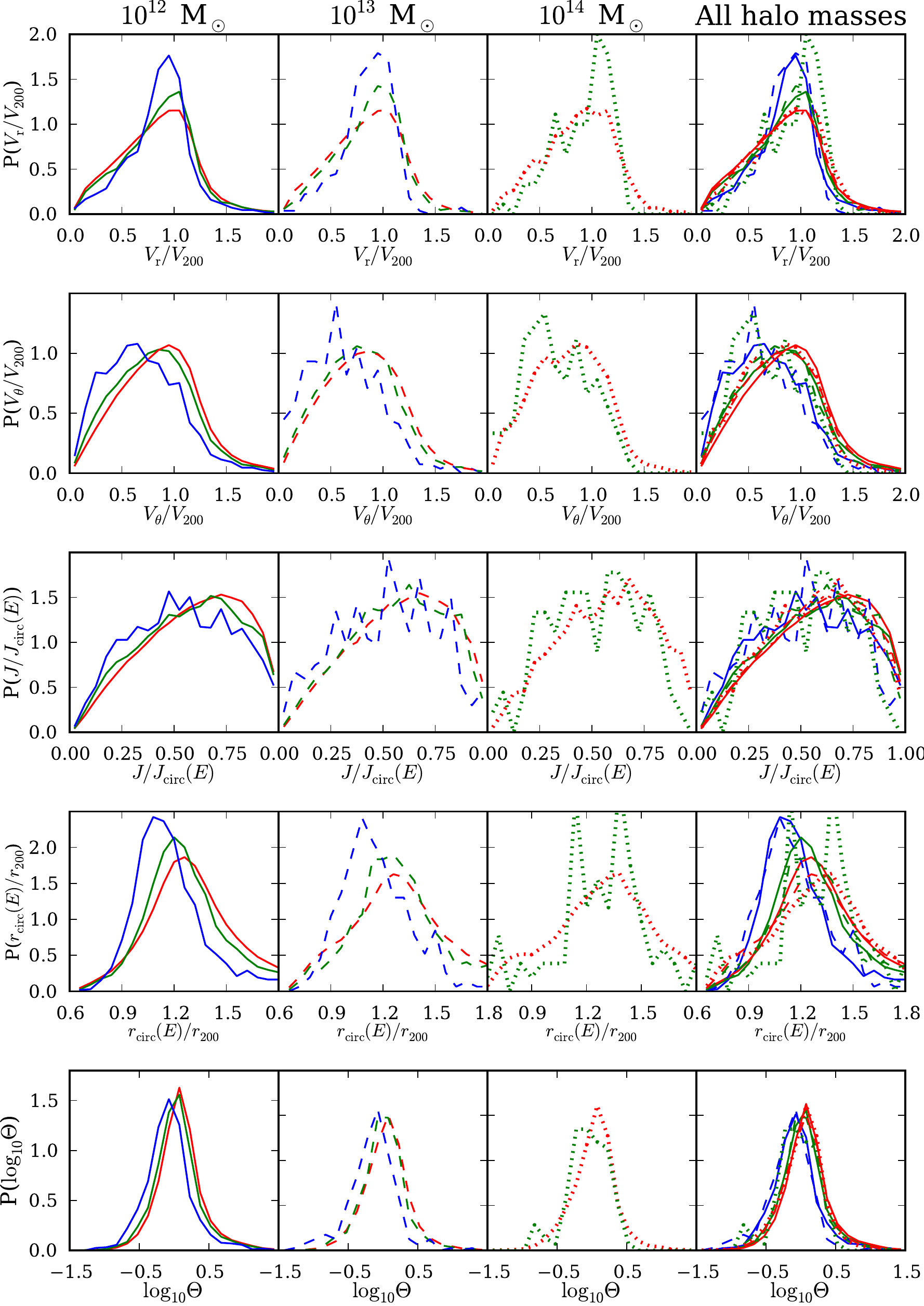}}%
\end{tabular}
\end{center}
\caption{Orbital parameter distributions for bins of different final
  halo masses and satellite-to-host halo mass ratios, $M_{\rm
    s}/M_{\rm h}$.  The central value of the final halo mass bin is
  indicated at the top of each column, with the rightmost column
  overplotting the results from each of the three mass bins using the
  appropriate line type.  The red lines are for $0.0001<M_{\rm
    s}/M_{\rm h}<0.005$, green for $0.005<M_{\rm s}/M_{\rm h}<0.05$
  and blue for $M_{\rm s}/M_{\rm h}>0.05$.  The first two rows show
  the radial, $V_r/V_{\rm 200}$, and tangential, $V_{\rm
    \theta}/V_{\rm 200}$, velocity distributions. The second two rows
  show the circularity, $J/J_{\rm circ}(E)$, and $r_{\rm
    circ}(E)/r_{200}$, while the final row shows the distributions of
  the composite parameter $\Theta$ defined in Eqn.~\ref{eq:Theta}.
  Note that for host haloes in the 10$^{14}$~M$_\odot$ bin, we do not
  show the $M_{\rm s}/M_{\rm h}>0.05$ distributions due to the low
  number of subhaloes.}
\label{fig:fig3}
\end{figure*}
 
The selection of \cite{tormen97} data which we plot matches the
$M_{\rm s}/M_{\rm h}>0.05$\footnote{We were able to apply this cut as
  G. Tormen kindly supplied his catalogue of satellite orbital
  parameters in electronic form.} cut used in our own data, but is for
host halos with typical masses of $10^{15}$~M$_\odot$. The good
agreement we find with \cite{tormen97} is only expected if, as we find
below, the distributions depend only weakly on halo mass at fixed
$M_{\rm s}/M_{\rm h}$. The \cite{benson05} data is based on a wide
range of simulations of different volumes and resolutions.  In this
sample he uses all satellites and haloes with masses greater than
$10^{11}$~M$_{\odot}$ and states that the typical ratio $M_{\rm
  s}/M_{\rm h}=0.08$.  The smooth radial and tangential velocity
distributions we plot in the lower panels of
Fig.~\ref{fig:fig_compare} are the fitted distributions presented by
\cite{benson05}. \cite{benson05} and also \cite{vit02} modelled the
radial distribution as a Gaussian and the tangential distribution as a
Rayleigh or 2D Maxwell-Boltzmann distribution. The agreement with our
results is reasonable.  The radial and tangential velocity
distributions of \cite{wetzel11} are in very good agreement with our
results. Like Benson, Wetzel uses all satellites and haloes above a
fixed mass cut, $10^{10}$~M$_{\odot}$, and so we would expect the mean
$M_{\rm s}/M_{\rm h}$ ratio to be similar to that of Benson and to our
$0.05<M_{\rm s}/M_{\rm h}<0.5$ sample.  The comparison of $J/J_{\rm
  circ}(E)$ distributions between us and Wetzel is not strictly fair
as we compute $J_{\rm circ}(E)$ using the singular isothermal sphere
model while he models the satellite and host as two point
masses. However while this introduces a bias for satellites for which
$M_{\rm s}/M_{\rm h}\ll 1$, we find that the resulting distributions
are very similar for satellites with $0.05<M_{\rm s}/M_{\rm h}<0.5$
(see Appendix~\ref{app:sis}).

\begin{figure*}
\resizebox{17cm}{!}{\includegraphics{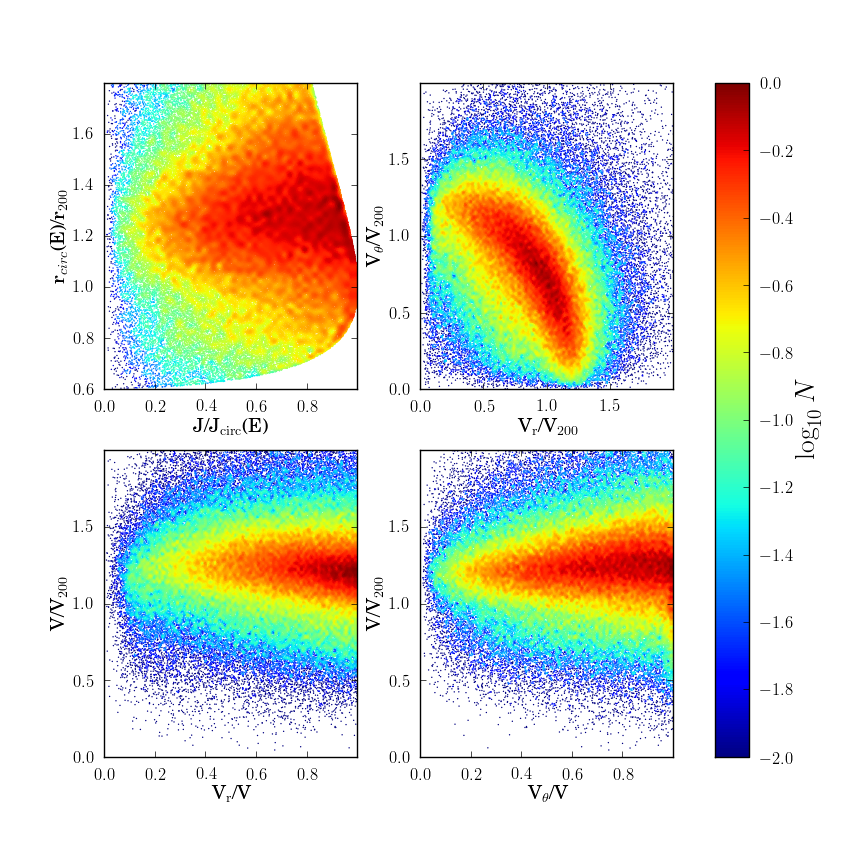}}\\%
\caption{The bivariate distributions of orbital parameters for all
  satellites infalling onto $10^{13}$~M$_\odot$ haloes. The top panels
  show the two-dimensional distribution of $r_{\rm circ}(E)/r_{\rm
    200}$ versus $J/J_{\rm circ}(E)$ and $V_{\rm \theta}/V_{\rm 200}$
  versus $V_{\rm r}/V_{\rm 200}$ respectively. The bottom panels show
  the two-dimensional distributions of $V/V_{\rm 200}$ versus
  $V_r/V_{\rm 200}$ and $V/V_{\rm 200}$ versus $V_{\theta}/V_{\rm
    200}$. The colour bar illustrates the relative density of points (on an
arbitrary scale). }
\label{fig:co}
\end{figure*}

\begin{table*}
  \caption{Parameters of the fitted orbital parameter distributions
    for bins of final halo mass, $M_{\rm h}$, and the satellite-to-host 
    mass ratio at infall, $M_{\rm s}/ M_{\rm h}$. The notation for the
    parameters of the Voigt and exponential fitting functions are
    as defined in Eqns.~\ref{eq:voigt} and \ref{eq:exp}.}
\label{table:table1}
\begin{center}
\begin{tabular}{|c |c |c |c |c |c |}
  \hline
  M$_{\rm h }$ &M$_{\rm s}$/ M$_{\rm h}$ & $B$& $\gamma$&$\sigma$  &
$\mu$ \hskip 1cm \\ \hline
10$^{12}$ M$_{\rm \odot }$
&$0.0001-0.005$&$0.049\pm{0.055}$&$\hphantom{-}0.109\pm{0.003}$&
$0.077\pm{0.002}$&$1.220\pm{0.001}$\\
10$^{13}$ M$_{\rm \odot}$
&$0.0001-0.005$&$0.548\pm{0.105}$&$\hphantom{-}0.114\pm{0.010}$&
$0.094\pm{0.006}$&$1.231\pm{0.002}$ \\
10$^{14}$ M$_{\rm \odot}$ &$0.0001-0.005$&$1.229\pm{0.292}$
&$\hphantom{-}0.110\pm{0.018}$&$0.072\pm{0.007}$&$1.254\pm{0.010}$ \\ \hline
10$^{12}$ M$_{\rm \odot}$&$0.005-0.05$ &$1.044\pm{0.086}$ &
$\hphantom{-}0.098\pm{0.005}$&$0.073\pm{0.004}$&$1.181\pm{0.002}$  \\
10$^{13}$ M$_{\rm \odot}$&$0.005-0.05$& $1.535\pm{0.255}$&
$\hphantom{-}0.087\pm{0.013}$&$0.083\pm{0.010}$&$1.201\pm{0.005}$\\
10$^{14}$ M$_{\rm \odot}$&$0.005-0.05$& $3.396\pm{1.040}$ &
$\hphantom{-}0.050\pm{0.023}$&$0.118\pm{0.025}$&$1.236\pm{0.020}$\\\hline
10$^{12}$ M$_{\rm
\odot}$&$0.05-0.5$&$2.878\pm{0.200}$&$\hphantom{-}0.071\pm{0.010}$&$0.091\pm{0.007}$&$
1.100\pm{0.004}$ \\
10$^{13}$ M$_{\rm \odot}$&$0.05-0.5$&
$3.946\pm{0.578}$&$\hphantom{-}0.030\pm{0.030}$&$0.139\pm{0.021}$&$1.100\pm{0.013}$ \\
10$^{14}$ M$_{\rm
\odot}$&$0.05-0.5$&$2.982\pm{4.646}$&$-0.012\pm{0.035}$&$0.187\pm{0.019}$&
$1.084\pm{0.052}
$   \\\hline
\end{tabular}
\end{center}
\end{table*}

\subsection{Orbital parameters: mass ratio and mass dependence}
\label{sub:mass}

Fig~\ref{fig:fig3} presents our results for the orbital parameter
distributions for three bins of halo mass and three bins of
satellite-to-host halo mass ratio.  We reiterate that the host halo
mass bins are defined by the mass of the host haloes at $z=0$ while
the mass ratio, $M_{\rm s}/M_{\rm h}$, is defined by the values at the
infall redshift.

The top two rows of Fig.~\ref{fig:fig3} show the distributions of
radial and tangential velocities at infall.  The radial distributions
peak close $V_{\rm r}=V_{200}$ and the tangential distributions 
at a lower value of around $V_\theta=0.65\, V_{200}$. 
As very few satellites are on
  unbound orbits, both distributions only have small
tails beyond $1.5\, V_{200}$.  Independently of host halo mass, we see
that the distributions of radial velocities become broader for lower
mass satellites with little change in the location of the peak of the
distribution.  In contrast for the tangential velocities the mode of
the distribution shifts to higher values for less massive
satellites. The most massive satellites are on the most radial, low
angular momentum, orbits, The dependence of these distributions on
halo mass at fixed $M_{\rm s}/M_{\rm h}$ is much weaker. This can be
seen in the righthand panels where, to a first approximation, the
lines of the same colour (same $M_{\rm s}/M_{\rm h}$) coincide.  There
is some residual dependence on halo mass (different line styles), with
orbits becoming more radial -- the $V_\theta/V_{200}$ distributions
peaking at lower values -- for more massive haloes, but this trend is
much weaker.

The middle row of Fig.~\ref{fig:fig3} shows the distributions
of circularity, $J/J_{\rm circ}(E)$. The distributions are broad
with those for the $M_{\rm s}/M_{\rm h}>0.05$ bin peaking
at close to a circularity of a half. In each bin of halo mass,
we again see the trend, for higher mass satellites to have less 
circular, more radially biased orbits. Also, once again,
 the trends with satellite-to-halo mass ratio are
much stronger than those with halo mass.

The penultimate row of Fig.~\ref{fig:fig3} shows the distributions of
$r_{\rm circ}(E)/r_{200}$. This is essentially a measure of the
energies of the orbits, with higher $r_{\rm circ}(E)/r_{200}$
corresponding to less bound orbits. At each halo mass, there is a
strong trend for the more massive satellites to be more strongly
bound. Again, the variations of the distributions with halo mass, at
fixed satellite-to-halo mass ratio, are much weaker.

These trends are consistent with the picture put forward by 
\cite{Libeskind05} that within the
filaments of the cosmic web that surround an accreting dark matter
halo, the most massive infalling haloes move along the central spines
of the filaments. In this way the filamentary structures act as
focusing rails which direct massive satellites onto predominantly
radial orbits.  Perhaps more simply, the force on the most massive
satellites is dominated by the central halo while lower mass
satellites can be significantly perturbed by other more massive
satellites.

We show the distribution of the composite orbital parameter
$\Theta$ in the bottom  row of Fig.~\ref{fig:fig3}
and list their median values in Table~\ref{table}. We 
see a clear shift in the distributions towards higher values of
$\Theta$ with decreasing values of $M_{\rm s}/M_{\rm h}$
and weaker dependence on host halo mass. According to
Eqn.~\ref{eq:taumrg} this will contribute to lower mass
satellites having longer merger timescales but this effect is
subdominant to the explicit $M_{\rm h}/M_{\rm s}$ term in that
equation which also acts in the same sense.

\subsection{2D distribution of orbital parameters}
\label{sub:2d}

As described in the \cite{benson05} paper, the radial and tangential
velocity distributions are tightly correlated. Consequently the
1-dimensional distributions presented in Fig.~\ref{fig:fig3} are not a
sufficient characterisation of the orbital parameter distributions.
We emphasise this in Fig.~\ref{fig:co} which shows bivariate
distributions of various orbital parameter combinations.

The top left-hand panel of Fig.~\ref{fig:co} shows the bivariate
distribution of $r_{\rm circ}(E)/r_{200}$ and $J/J_{\rm circ}(E)$.  The
first thing to note in this distribution is that there are excluded
regions at high value of $J/J_{\rm circ}(E)$ both for low and high values
of $r_{\rm circ}(E)/r_{200}$.  These arise from our stipulation that
we are characterising the orbits of satellites when they first cross
$r_{200}$. The plotted distribution touches the right hand axis at
$r_{\rm circ}(E)/r_{200} =1$ and $J/J_{circ}(E)=1$. This point
corresponds to a circular orbit with $r=r_{200}$. Circular orbits of
either larger or smaller radius would not be included in our sample as
they never cross $r_{200}$. Hence, $r_{\rm circ}(E)/r_{200}$
  either increases of decreases away from unity for increasingly
  eccentric orbits in our sample. This defines the complicated
boundary to the measured bivariate distribution.

The top right-hand panel of Fig.~\ref{fig:co} shows the correlated
bivariate distribution of radial and tangential velocities. This is
similar to that presented and parametrised in \cite{benson05}.  We
note that the ridge line of this distribution is approximately
circular, i.e. it corresponds to a fixed total velocity $V=\left(V_r^2
  +V_\theta^2 \right)^{1/2}$.

The lower two panels of Fig.~\ref{fig:co} show the two dimensional
distributions of the total velocity versus either the ratio $V_r/V$ or
$V_\theta/V$.  We see to a good approximation these pairs of
parameters appear uncorrelated. This suggests that we can construct a
simple model for the full bivariate distribution of orbital parameters
by modelling the individual independent distributions of $V/V_{200}$
and $V_r/V$. This will then provide a simple parametrised model that
can be used in semi-analytic galaxy formation models.

\subsection{Fitted Distributions}
\label{sec:fits}

To build a complete model of the bivariate distribution of orbital
parameters we perform fits to the marginalised distributions of both
the total velocity, $V/V_{200},$ and the radial-to-total velocity
ratio, $V_r/V$. Assuming these to be independent we can then transform
variables to generate model predictions for the distributions of any
of the other choices of orbital parameters such as $J/J_{\rm circ}(E)$
and $r_{\rm circ}(E)/r_{200}$. Here we present these fits as a
function of halo mass and satellite-to-halo mass ratio.

\begin{figure}
\includegraphics[width=8.5cm]{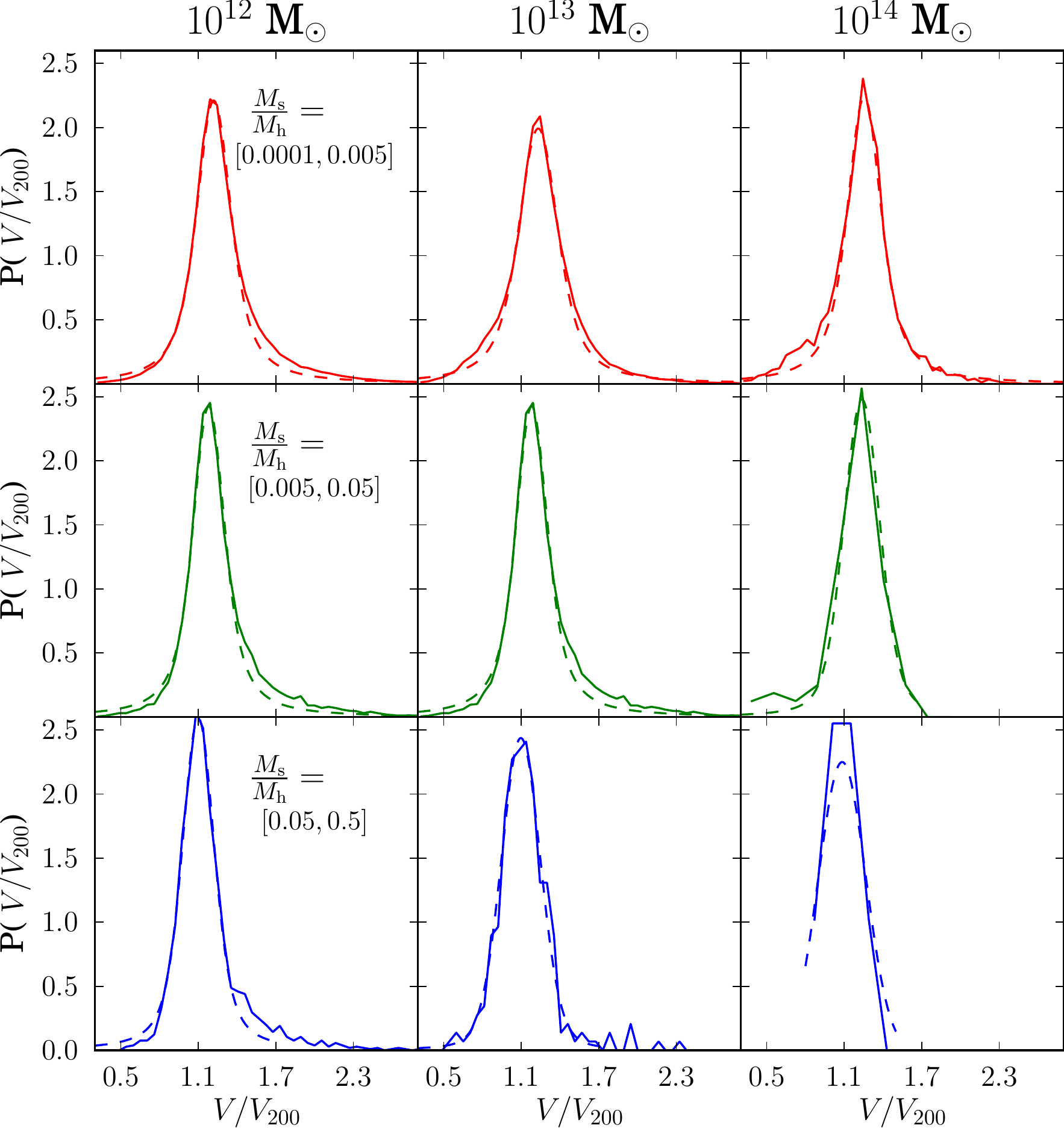}%
\caption{Probability distribution of the total infall velocity,
  $V/V_{200}$, as a function of both the satellite-to-host mass ratio
  at infall and the host halo mass. Each column is for a fixed final
  halo mass as labelled at the top of the column. Each row is for a
  different bin in satellite-to-host mass ratio: top (red lines)
  $0.0001<M_{\rm s}/M_{\rm h}<0.005$ , middle (green lines)
  $0.005<M_{\rm s}/M_{\rm h}<0.05$ and bottom (blue lines) $M_{\rm
    s}/M_{\rm h}>0.05$. The dashed lines are the Voigt profile fits
  whose parameters, $\mu$, $\gamma$ and $\sigma$ are listed in
  Table~\ref{table:table1}.  }
\label{fig:voigt}
\end{figure}

The distributions of $V/V_{200}$ for each of our samples are shown in
Fig.~\ref{fig:voigt} along with Voigt profile fits.  The distributions
of $V/V_{200}$ are reasonably symmetric about their means but much
more centrally peaked than Gaussians of the same rms width
(leptokurtic). We find that the distributions can be fitted well by
Voigt profiles, convolutions of a Lorentz profile,
\begin{equation}
P_L(x; \gamma) \equiv \frac{\gamma }{\pi(x^{2}+\gamma^{2})}, 
\end{equation} and a Gaussian 
\begin{equation}
P_G(x; \sigma,\mu) \equiv \frac{1}{\sqrt{2\pi} \, \sigma}
\exp \left( \frac{-(x-\mu)^{2}}{2\sigma^2} \right)
\end{equation}
\begin{equation}
P_V(x; \sigma , \gamma , \mu) = \int_{-\infty}^{+\infty}
P_G(x^{\prime};\sigma , \mu ) P_L(x-x^{\prime}; \gamma )dx^{\prime}
\label{eq:voigt}
\end{equation}
where $x=V/V_{200}$. We determine the best fitting Voigt profiles by
finding the parameters that maximise the likelihood, ${\cal L} = \Pi_i
P_V(x_i; \sigma , \gamma , \mu)$, where the index $i$ runs over all
the satellites in the sample.  The resulting fits are shown in
Fig.~\ref{fig:voigt} and their parameters $\sigma$, $\gamma$ and $\mu$
are listed in Table.~\ref{table:table1}.

\begin{figure}
\includegraphics[width=8.5cm]{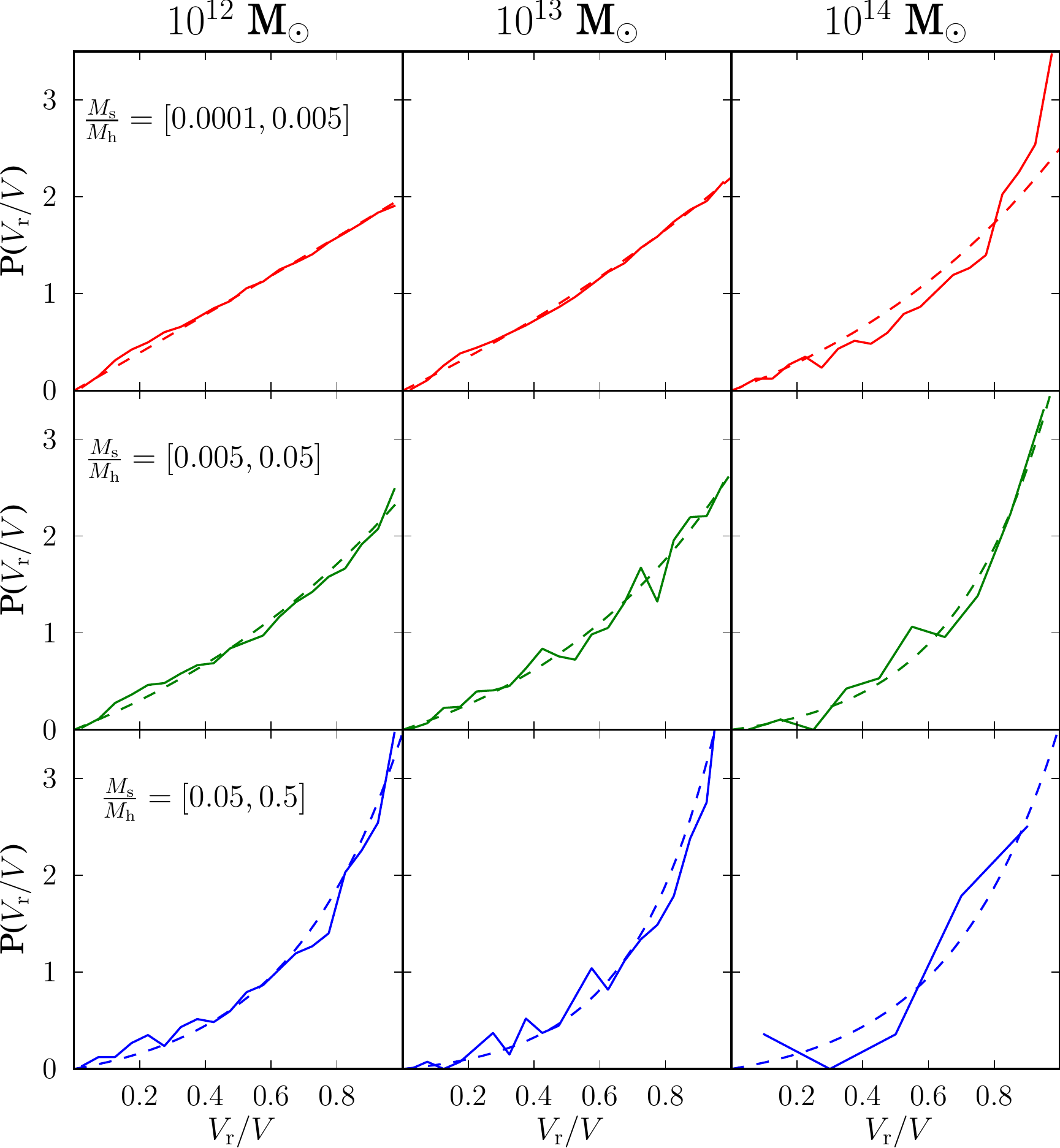}%
\caption{Dependence of the orbital parameters $V_r/V$
  on the mass ratio between the satellite halo mass and the host halo
  mass. 
  Each column is for a fixed final halo mass as labelled
  at the top of the figure. Each row is for a different bin
  in satellite-to-host mass ratio, top (red lines) 
  $0.0001<M_{\rm s}/M_{\rm h}<0.005$, 
  middle (green lines) $0.005<M_{\rm s}/M_{\rm h}<0.05$
  and bottom (blue lines)  $M_{\rm s}/M_{\rm h}>0.05$
  The dashed curves are the best fitting exponential distributions
  and the corresponding value of the parameter $B$ in Eqn.~\ref{eq:exp}) 
  is shown on each panel and in Table~\ref{table:table1}.
}
\label{fig:vrandv}
\end{figure}

\begin{figure}
\includegraphics[width=8cm]{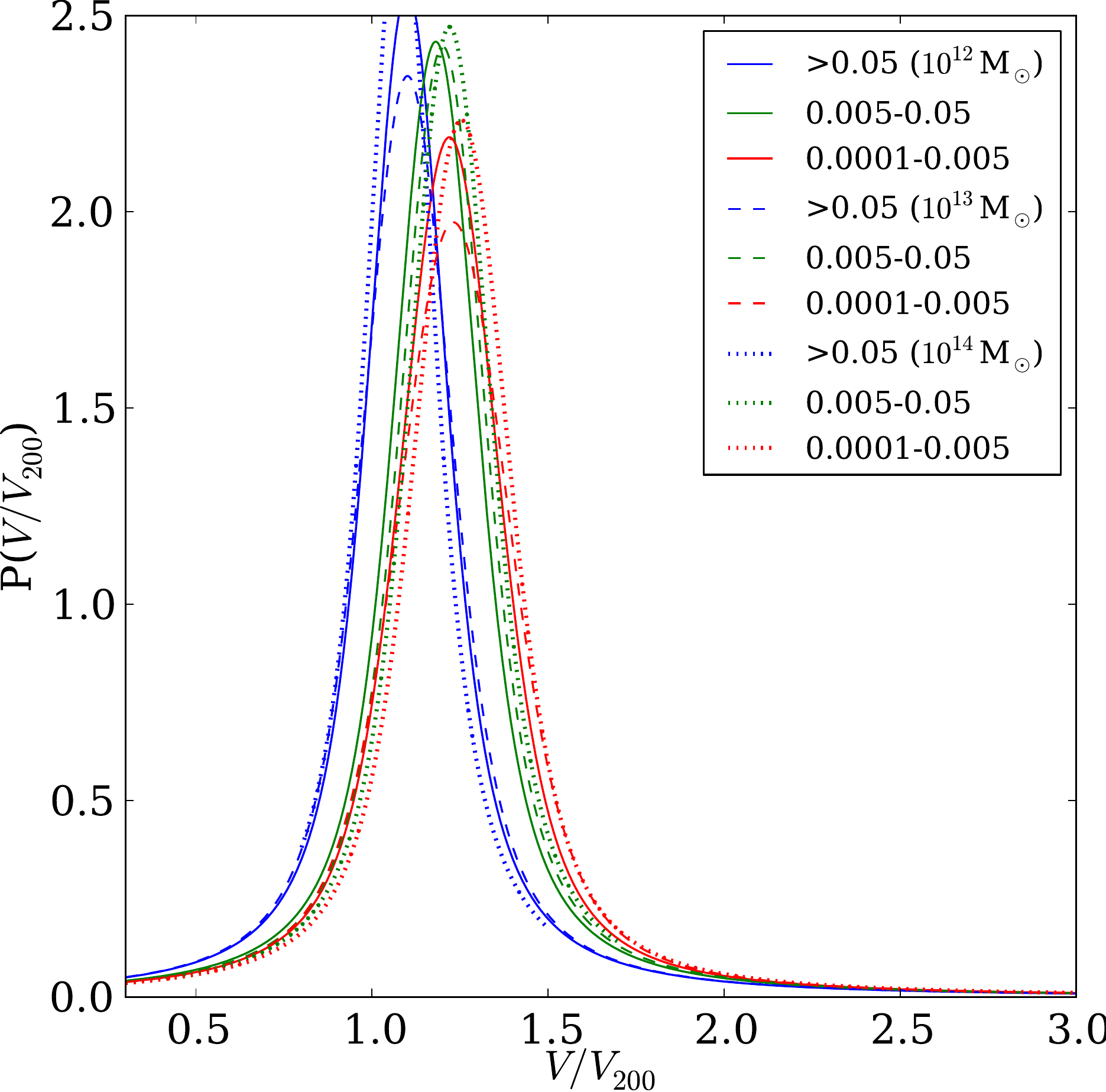}\\
\includegraphics[width=8cm]{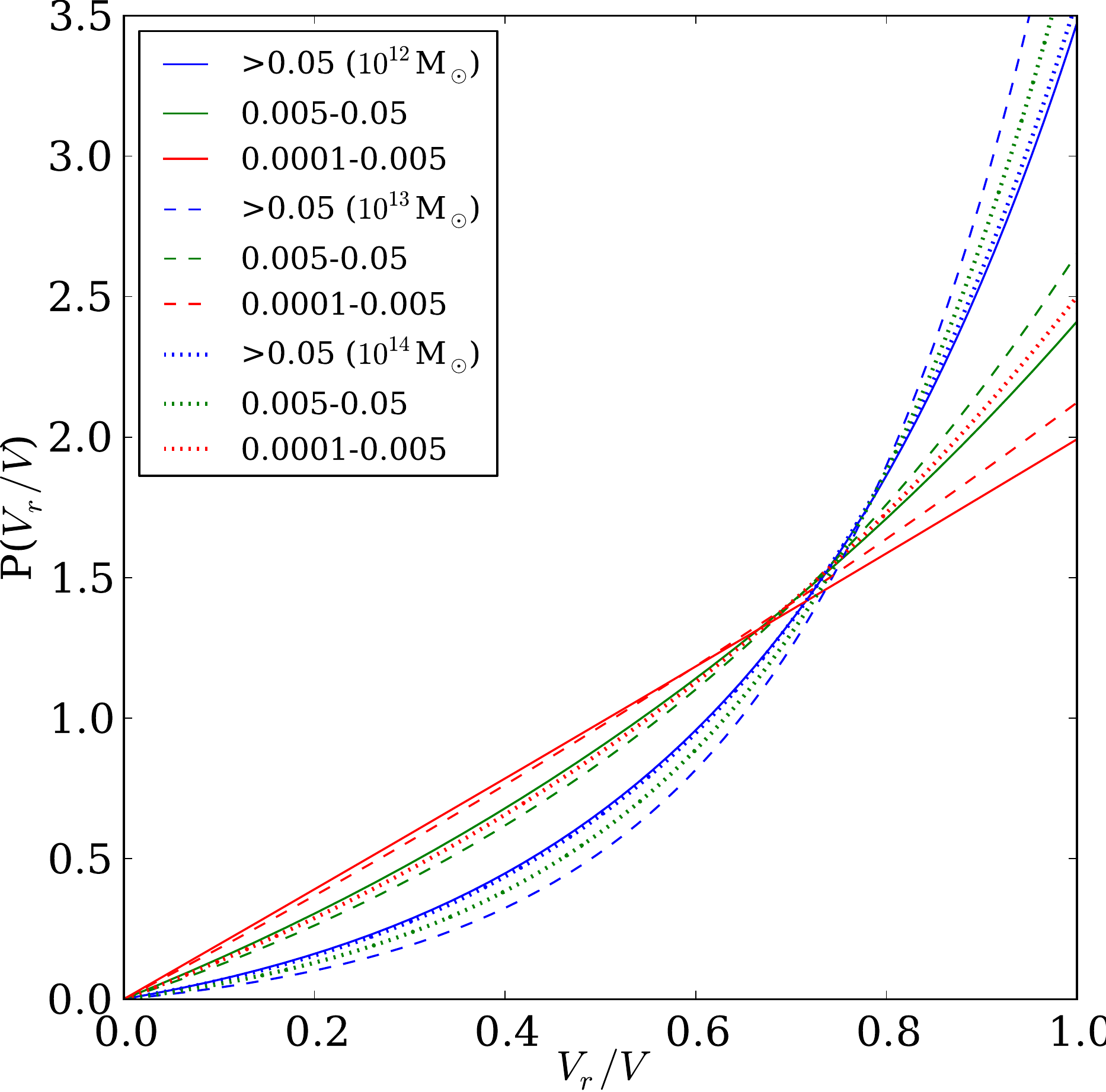}%
\caption{The fitted distributions of 
the orbital parameters $V/V_{200}$ (top) and $V_r/V$ (bottom)
  for the different values of both the satellite-to-host mass ratio
  and the host halo mass. Line colour denotes satellite-to-host mass
  ratio, red   $0.0001<M_{\rm s}/M_{\rm h}<0.005$, green 
  $0.005<M_{\rm s}/M_{\rm h}<0.05$
  and blue  $M_{\rm s}/M_{\rm h}>0.05$. The line style
  indicates the host halo mass, solid $10^{12}$~M$_\odot$,
  dashed $10^{13}$~M$_\odot$ and dotted $10^{14}$~M$_\odot$.
}
\label{fig:curve}
\end{figure}

\begin{figure}
\includegraphics[width=8.5cm]{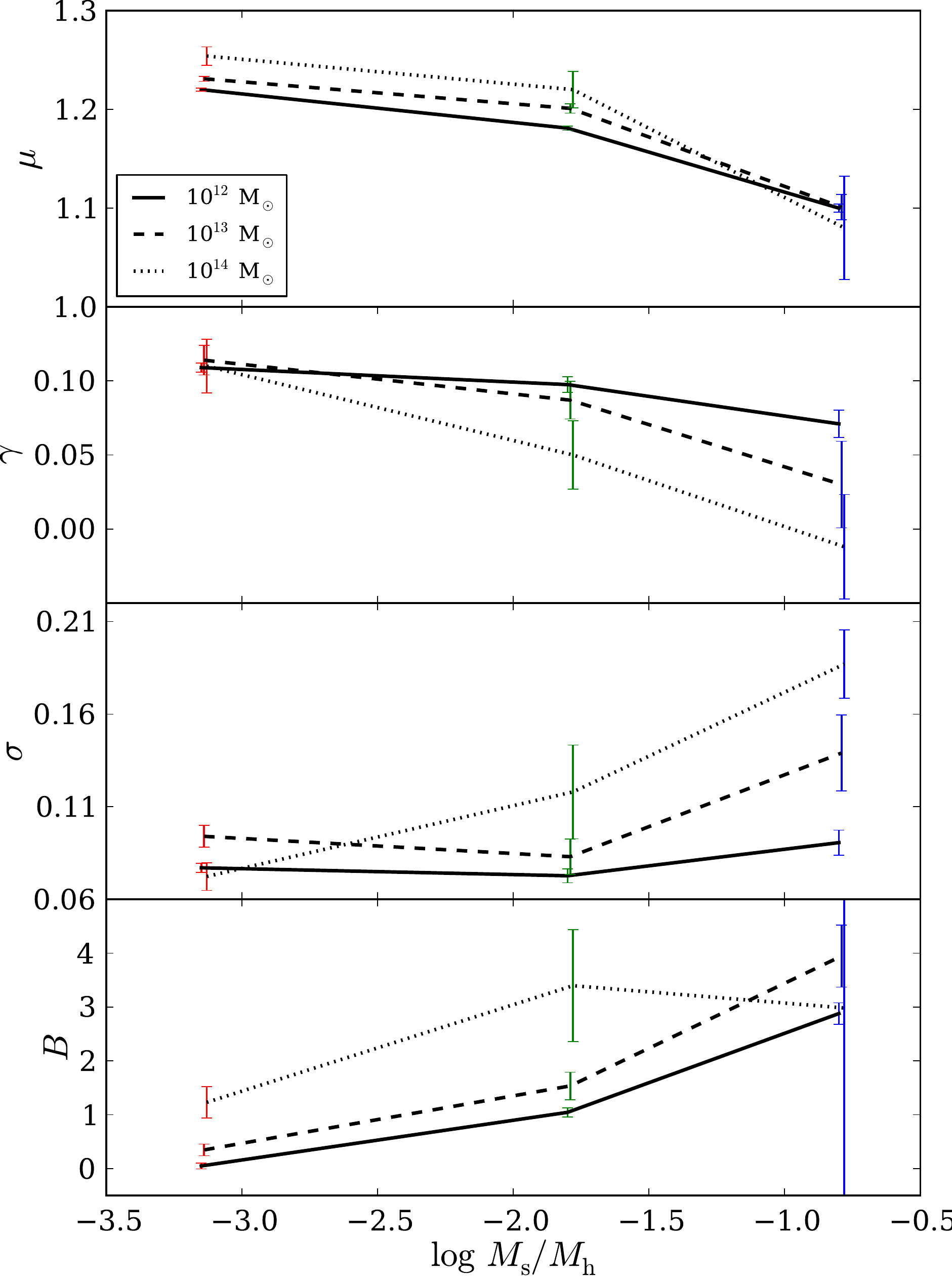}
\caption{Dependence of fitting parameters $\mu$, $\gamma$, $\sigma$
  and $B$ on the satellite-to-host mass ratio at infall.  Thetable
  different line styles denote different host halo masses, as
  indicated in the legend. The colours of the error bars denote
  satellite-to-host mass ratio bins: red $0.0001<M_{\rm s}/M_{\rm
    h}<0.005$, green $0.005<M_{\rm s}/M_{\rm h}<0.05$ and blue $M_{\rm
    s}/M_{\rm h}>0.05$. The errors are estimated by bootstrap sampling
  of the $z=0$ halo catalogue.}
\label{fig:fit}
\end{figure}

We find that the distributions $V_r/V$ are well fit by exponential
distributions of the form:
\begin{equation}
P(V_r/V) = A\left(\exp\left( \frac{BV_r}{V}\right)-1 \right).
\label{eq:exp}
\end{equation}
Here $A$ is simply a normalisation constant and $B$ is the single free
parameter. The distributions of $V_r/V$ and the corresponding maximum
likelihood fits are shown in Fig.~\ref{fig:vrandv}.  The distribution
is almost linear, $B\ll 1$, for the combination of low $M_{\rm h}$ and
low $M_{\rm s}/M_{\rm h}$. The distributions become increasingly
radially biased, peaked at $V_r/V=1$ (high $B$), for both increasing
$M_{\rm s}/M_{\rm h}$ and $M_{\rm h}$, consistent with our earlier
discussion.

The trends of the distributions of $V/V_{200}$ and $V_{\rm r}/V$ with halo
mass and satellite-to-halo mass ratio are depicted more clearly in
Fig.~\ref{fig:curve}, which shows all the fitted distributions on a
single panel. In the lower panel we see the tendency for the
distributions to become more radially biased for satellites with
higher $M_{\rm s}/M_{\rm h}$.  In the upper panel, it is clear that
the $V/V_{200}$ distributions have very little dependence on halo mass
at fixed $M_{\rm s}/M_{\rm h}$.  There is a stronger dependence on
$M_{\rm s}/M_{\rm h}$ with samples of larger $M_{\rm s}/M_{\rm h}$
ratios having narrower distributions and lower mean values. This is
consistent with the similar trends in the distribution of $r_{\rm
  circ}(E)/r_{200}$ that we saw in Fig.~\ref{fig:fig3}.  These trends
can also be seen in Fig.~\ref{fig:fit}, where we plot the dependence
of the fit parameters on $M_{\rm s}/M_{\rm h}$. In all halo mass bins
the mean, $\mu$, decreases strongly for the highest values of $M_{\rm
  s}/M_{\rm h}$. The narrower width of the $V/V_{200}$ distributions
for high $M_{\rm s}/M_{\rm h}$, which we see in Fig.~\ref{fig:curve},
is reflected in a decreasing value of $\gamma$ (the width of the
Lorentzian) with increasing $M_{\rm s}/M_{\rm h}$, which has greater
effect on the width of the distribution than the corresponding slow
increase in $\sigma$ (the width of the Gaussian).  The error bars
shown on Fig.~\ref{fig:fit} have been estimated by bootstrap
resampling of the $z=0$ halo catalogue and we have investigated the
correlations of all the pairs of parameters. The only significant
correlation we find is an anticorrelation between $\sigma$ and
$\gamma$. This is to be expected as the overall width of the
distribution is determined by $\sigma^2+\gamma^2$, while their ratio,
$\gamma/\sigma$, determines how peaked the distribution is (its
kurtosis).

\subsection{Derived Distributions}
\label{sec:derived}

\begin{figure*}
\begin{center}
\begin{tabular}{cc}
\resizebox{15.0cm}{!}{\includegraphics{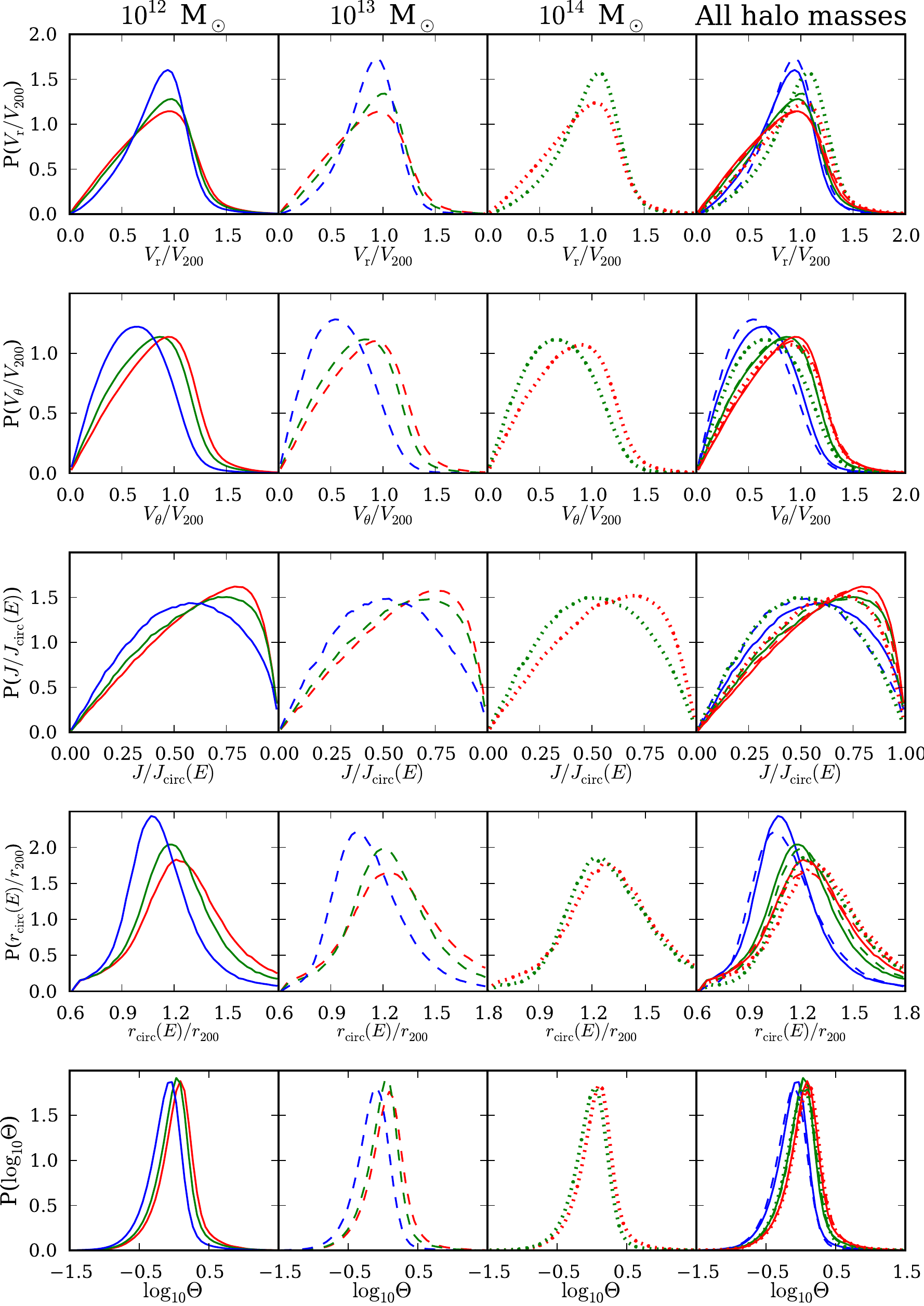}}%
\end{tabular}
\end{center}
\caption{Like Fig.~\ref{fig:fig3}, but showing the distributions
derived from the fits presented in Section~\ref{sec:fits}
rather than the directly measured distributions.}
\label{fig:fig3fit}
\end{figure*}

If the fits we have presented in Section~\ref{sec:fits} are accurate
and if $V_r/V$ and $V/V_{\rm 200}$ are uncorrelated then we can use
these distributions to derive model distributions of any other choice
of orbital parameter. For instance we can select pairs of values of
$V_r/V$ and $V/V_{\rm 200}$ from the fitted
distributions and compute the radial and tangential velocities using
\begin{equation}
\frac{V_r}{V_{200}} = \left(\frac{V_r}{V}\right) 
\left( \frac{V}{V_{200}} \right)
\label{equation:vr}
\end{equation}
and
\begin{equation}
\frac{V_\theta}{V_{200}} = \left( \frac{V}{V_{200}} \right)
\sqrt{1- \left(\frac{V_r}{V}\right)^2 }.
\label{equation:vtan}
\end{equation}
We can also derive $J/J_{\rm circ}(E)$, $r_{\rm circ}(E)/r_{200}$ and
$\Theta$ from $V_r/V$ and $V/V_{\rm 200}$ using the equations in
Section~\ref{sub:method}. We show all the resulting orbital parameter
distributions in Fig.~\ref{fig:fig3fit}, which should be compared with
Fig.~\ref{fig:fig3}. Direct comparison of the two figures shows that
these are faithful representations of the data and validate the
assumption that, to a good approximation, $V_r/V$ and $V/V_{\rm 200}$
can be treated as independent random variables. The model
distributions shown in Fig.~\ref{fig:fig3fit}, particularly the
superimposed distributions in the righthand column, clearly show both
the strong dependence on $M_{\rm s}/M_{\rm h}$ and the much weaker
dependence on $M_{\rm h}$.

The values of $V$ and $V_{\rm r}$ are directly measured and so the
fitted distributions of $V_r/V$ and $V/V_{\rm 200}$ make no assumption
about the form of the density profile of the host halo. Hence if 
prefered one  one can compute
the corresponding orbital parameters 
$r_{\rm circ}(E)/r_{200}$ and  $J/J_{\rm circ}(E)$ using an
NFW model of the halo profile. Following the same steps as outlined in
Section~\ref{sub:method} but for an NFW profile one can show that
$r_{\rm circ}(E)/r_{200}$ can be evaluated by solving 
\begin{equation}
\frac{r_{\rm circ}(E)}{r_{200}}
= \frac{2\ln(1+c\, r_{\rm circ}(E)/r_{200})-1/f(c\, r_{\rm circ}(E)/r_{200})}{2\ln(1+c)-V^2/(V_{200}^2 f(c))}
\end{equation}
where $f(c)=1/(\ln(1+c)-c/(c+1))$ and $c$ is the
concentration. \footnote{This equation can be solved iteratively by starting with
the guess $r_{\rm circ}(E)/r_{200}=1$ and obtaining the next iteration by
substituting into the RHS.} Having found $r_c(E)/r_{200}$,
$J/J_{\rm circ}(E)$ can be found using
\begin{equation}
\frac{J}{J_{\rm circ}(E)} =\frac{V_\theta}{V_{200}} 
\left( \frac{f(c\, r_{\rm circ}(E)/r_{200})}{f(c)} \right)^{1/2}
\left(\frac{r_{\rm circ}(E)}{r_{200}}\right)^{-1/2} .
\end{equation}

\begin{figure*}
\resizebox{7.5cm}{!}{\includegraphics{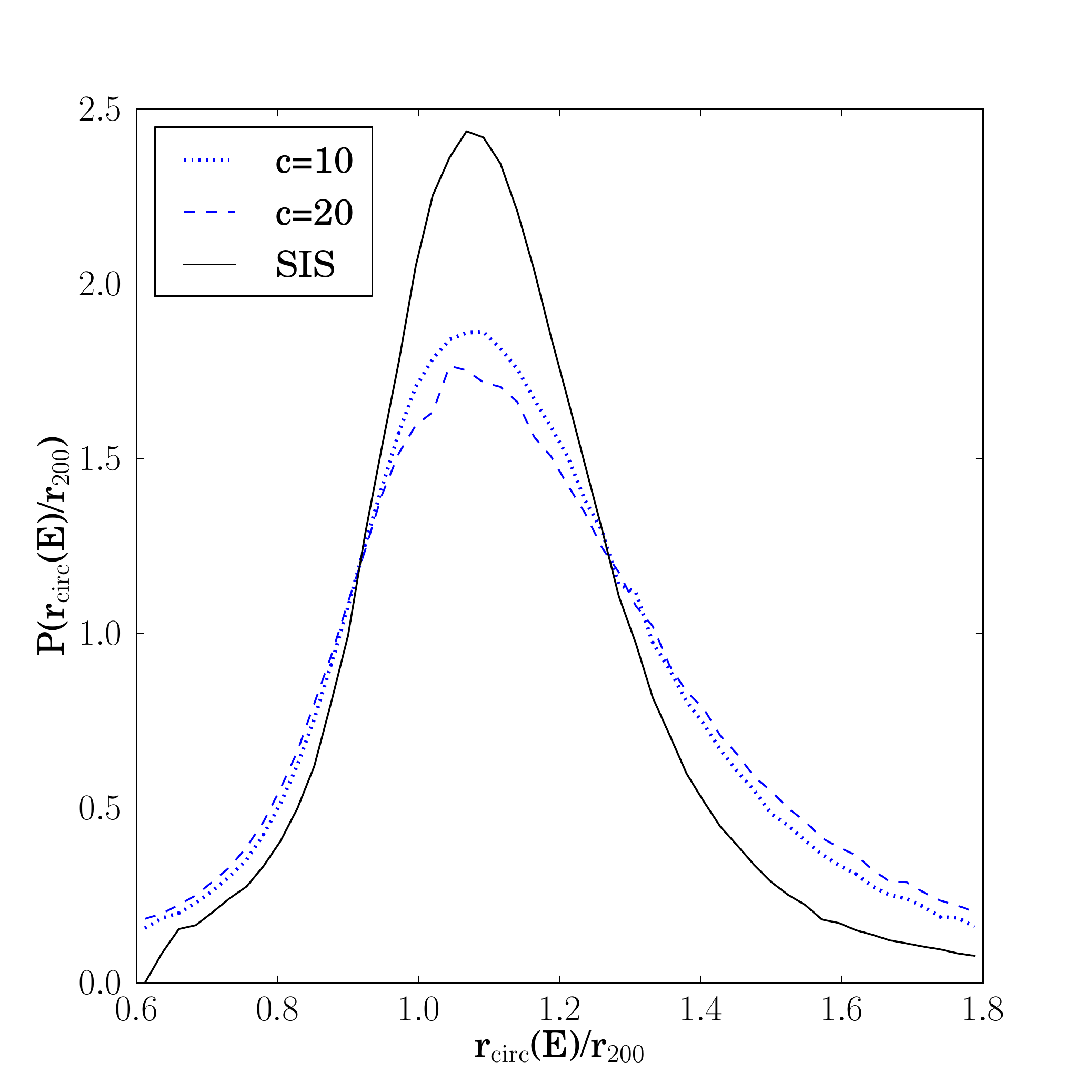}}
\resizebox{7.5cm}{!}{\includegraphics{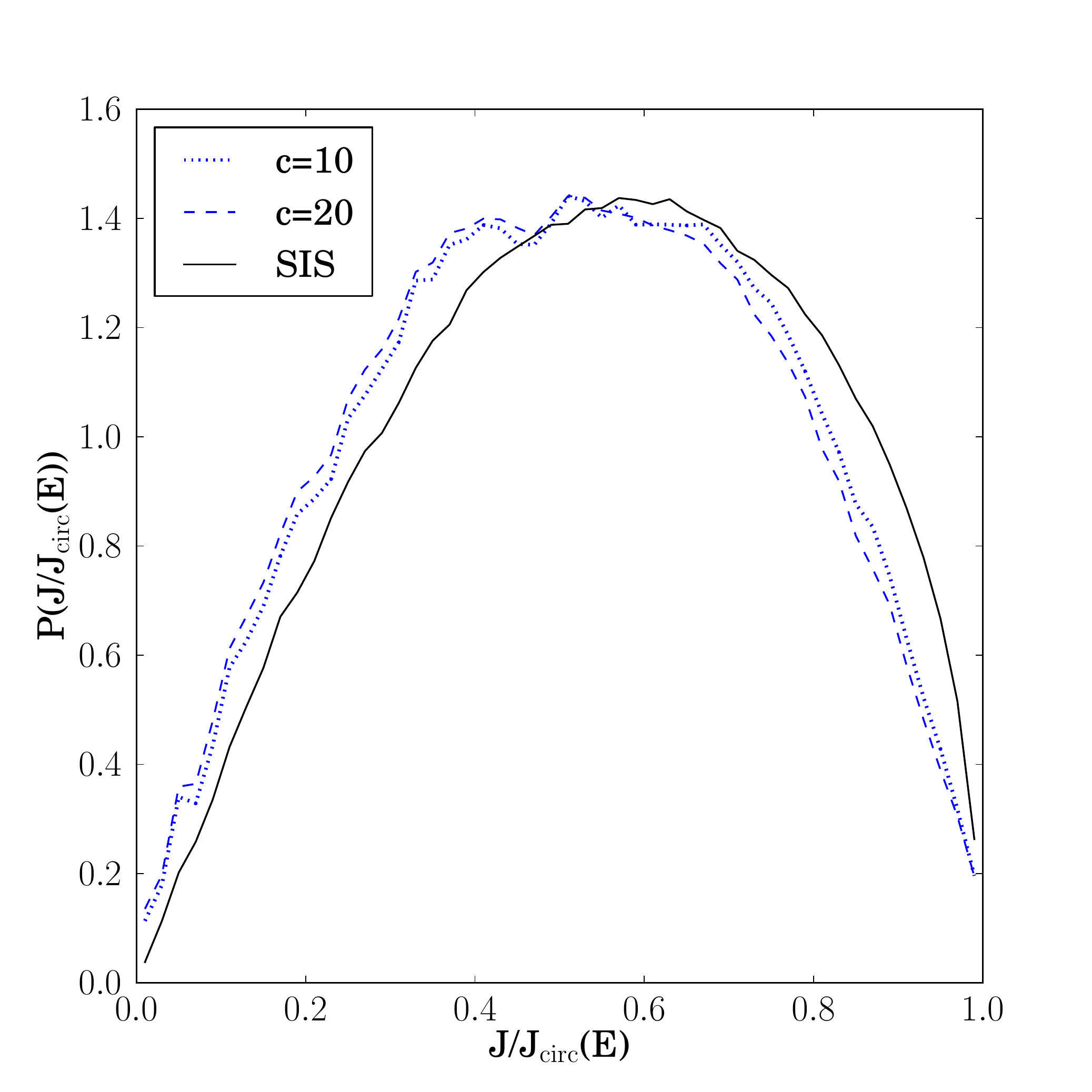}}
\caption{
Comparison of the fitted distributions of orbital parameters
$r_{\rm circ}(E)/r_{200}$, $J/J_{\rm circ}(E)$ for different
assumed host halo density profiles. The black solid curve is for
the default SIS profile when the blue dotted and dashed curves are for
NFW profiles with concentrations $c=$10 and~20 respectively.
This example is for host halo mass $M_{\rm h}=10^{12}$~M$_\odot$
and satellite to mass ratio in the range 
$0.05<M_{\rm s}/M_{\rm h}<0.5$.
 }
 \label{fig:fig_NFW}
\end{figure*}
As an example, Fig.~\ref{fig:fig_NFW} compares the resulting
distributions of $r_{\rm circ}(E)/r_{200}$ and 
$J/J_{\rm circ}(E)$ 
for host halo mass $M_{\rm h}=10^{12}$~M$_\odot$
and satellite to mass ratio in the range 
$0.05<M_{\rm s}/M_{\rm h}<0.5$. The NFW distributions
of $r_{\rm circ}(E)/r_{200}$ are broader than the corresponding
SIS distribution as over the relevant range 
$r_{\rm circ}(E)$ is a stronger function of energy in the NFW case than it
is in the SIS case. The energy of a circular orbit at $r=r_{200}$ 
is the same for both NFW and SIS as both profiles are normalized to
enclose the same mass, $M_{200}$, at this radius. As the orbital energy
is increased $r_{\rm circ}(E)$ will grow faster for the NFW case as
it has a more rapidly decreasing density profile. 
This leads to the enhanced tail
of orbits with large $r_{\rm circ}(E)/r_{200}$. Higher concentration
steepens the density profile around $r_{200}$ and so further enhances
this tail, but this is a very weak effect. The 
$J/J_{\rm circ}$ distributions for the NFW models are slightly shifted
to lower values. Since the measured values of $J$ do not depend on the
model this shift is caused by the typical values of $J_{\rm circ}$
being larger in the NFW case. For a circular orbit at $r_{200}$ the
NFW and SIS models have identical  $J_{\rm circ}$ while at larger
radii $J_{\rm circ}$ for a NFW prfile exceeds that for a SIS. Hence
the small shift to lower $J/J_{\rm circ}$ is a result of this sample
having a median value of $r_{\rm circ}(E)/r_{200}$ greater than unity.
The shift is larger for lower mass satellites as their distributions of
 $r_{\rm circ}(E)/r_{200}$ have larger median values.
The dependence on concentration is again very weak.

\section{Conclusions}

We have employed the DOVE high resolution cosmological N-body simulation
with more than 4 billion particles to study the distribution of the orbits
of infalling satellites during hierarchical halo formation
in the standard $\Lambda$CDM cosmology.
We study host haloes with masses from $10^{12}$ to $10^{14}$~M$_\odot$
and satellites with masses as low as $2\times10^8$~M$_\odot$.
Compared to previous studies \citep{tormen97, vit02, benson05, wetzel11}
we have better mass and time resolution and a larger sample
of satellite orbits.

There are various choices for the pair of orbital parameters that
specify a satellite orbit in a spherical potential. We quantify the
distributions of the radial, $V_{\rm r}$, and tangential, $V_\theta$,
velocities as well as other common alternatives such as the
circularity, $J/J_{\rm circ}(E)$ and the radius of the circular orbit
of the same energy, $r_{\rm circ}(E)$.

We have examined the dependence of the distributions of these orbital
parameters on both the host halo mass, $M_{\rm h}$, and the mass ratio
between the satellite and host, $M_{\rm s}/M_{\rm h}$.  We find that
the strongest trends are with $M_{\rm s}/M_{\rm h}$ at fixed $M_{\rm
  h}$. Satellites with larger $M_{\rm s}/M_{\rm h}$ tend to be on more
radial orbits with lower angular momentum and are more tightly bound.
At fixed $M_{\rm s}/M_{\rm h}$ there is a trend for satellites around
more massive haloes to also be on more radial orbits, but this trend
is weaker. Insofar as previous authors have examined similar
  relationships, our results are consistent with their data. However,
  while \cite{wetzel11} had not detected a significant dependence
  of orbital parameters on satellite mass ratio, possibly due to their
  limited sample size, our larger sample of orbits reveals a
  dependence, particularly at high mass ratios.

In general we find that complementary pairs of orbital parameters,
such as ($V_{\rm r}$,$V_\theta$), are non-trivially correlated, making
a complete description of their bivariate distribution
complex. However, we find that, to a good approximation, the
distributions of total infall velocity $V=(V_{\rm
  r}^2+V_\theta^2)^{1/2}$ and the ratio $V_r/V$ are uncorrelated. We
present accurate Voigt and exponential fits to their respective
distributions.  Assuming them to be uncorrelated, we transform these
simple bivariate distributions and demonstrate that the distributions
of other choices of orbital parameter can be successfully recovered.

\section*{acknowledgements}

This work was supported by the Science and Technology Facilities 
[ST/L00075X/1]. LJ acknowledges the support of a Durham Doctoral
Studentship and CSF an ERC Advanced Investigator grant Cosmiway [GA
267291]. This work used the DiRAC Data Centric system at Durham
University, operated by the Institute for Computational Cosmology on
behalf of the STFC DiRAC HPC Facility (www.dirac.ac.uk). This equipment
was funded by BIS National E-infrastructure capital grant ST/K00042X/1,
STFC capital grant ST/H008519/1, and STFC DiRAC Operations grant
ST/K003267/1 and Durham University. DiRAC is part of the National
E-Infrastructure.

\bibliographystyle{mn2e}

\bibliography{rev}

\appendix
\section{Circularity in the Keplerian Approximation}
\label{app:sis}

To compare the circularity, $J/J_{\rm circ}(E)$, inferred under the
assumption that the infalling satellite and host halo are treated as
two point masses on a Keplerian orbit with the singular isothermal
sphere (SIS) model we need to compare the corresponding expressions
for the angular momenta of circular orbits, $J_{\rm circ}(E)$. For the
Keplerian case this is easily derived from the angular momentum of a
circular orbit of radius $r$, $J_{\rm circ} = \mu V_{\rm circ} r$,
where the circular velocity at separation $r$ is given by $\mu V_{\rm
  circ}^2= G M_{\rm h} M_{\rm s}/r $ and the corresponding orbital
energy $E= \mu V_{\rm circ}^2/2 - G M_{\rm h} M_{\rm s}/r$.  Here
$\mu$ is the reduced mass, which can be expressed in terms of the
satellite and host masses as $\mu = M_{\rm s}M_{\rm h}/\left(M_{\rm
    s}+M_{\rm h}\right)$.  Eliminating both $V_{\rm circ}$ and $r$
from these three equations yields
\begin{equation}
J^{\rm Kep}_{\rm circ}(E) = \sqrt{\frac{(G M_{\rm h} M_{\rm s})^2
    \mu}{-2E}}.
\label{eq:J_Kep}
\end{equation}
If $V$ is the velocity difference between the satellite and host when the
satellite crosses the virial radius, $r_{200}$, then
\begin{equation}
E=\frac{1}{2}\mu V^2-\frac{GM_{\rm s}M_{\rm h}}{r_{200}} 
=\frac{1}{2}\mu V^2-M_{\rm s}V_{200}^2 ,
\label{eq:E_kep}
\end{equation}
where the circular velocity, $V_{200}$, is given by
$V_{200}=\sqrt{GM_{\rm h}/r_{200}}$.  Using Eqn.~\ref{eq:E_kep} to
substitute for $E$ in Eqn.~\ref{eq:J_Kep} yields
\begin{equation}
\frac{ J^{\rm Kep}_{\rm circ}(E)}
{   M_{\rm s} V_{200}\, r_{200}}
=\frac{1}{\sqrt{2M_{\rm s}/\mu - V^2/V_{200}^2 }} .
\end{equation}
This compares with the singular isothermal sphere expression for
$J_{\rm circ}$ derived in Section~\ref{sub:method},
\begin{equation}
\frac{ J^{\rm SIS}_{\rm circ}(E)}
{   M_{\rm s} V_{200}\, r_{200}} 
= \exp\left(\frac{1}{2}
\left( \frac{V^2}{V_{200}^2} - 1 \right)\right) .
\end{equation}

\begin{figure}
\includegraphics[width=10cm]{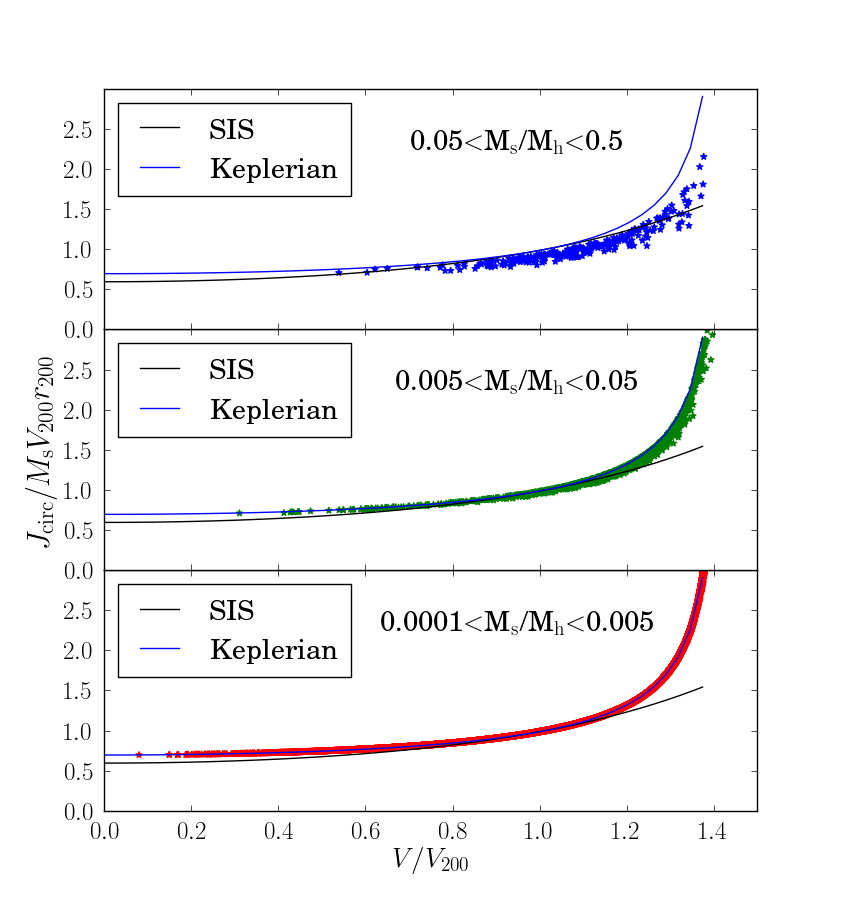}
\caption{A comparison of the Keplerian and singular isothermal sphere
  (SIS) models of $J_{\rm circ}$ in units of $M_{\rm s} V_{200}\,
  r_{200}$ for satellites with infall velocity, $V$, at the virial
  radius $r_{200}$.  In each panel, the black solid line is the SIS
  expression and the blue solid line is for the Keplerian case in the
  limit $M_{\rm s}/M_{\rm h}\ll 1$. The stars show the result of the
  full Keplerian expression including the dependence on the reduced
  mass, $\mu$, for for samples of satellites in different bins of
  $M_{\rm s}/M_{\rm h}$.  }
\label{fig:A1}
\end{figure}

\begin{figure}
\includegraphics[width=8cm]{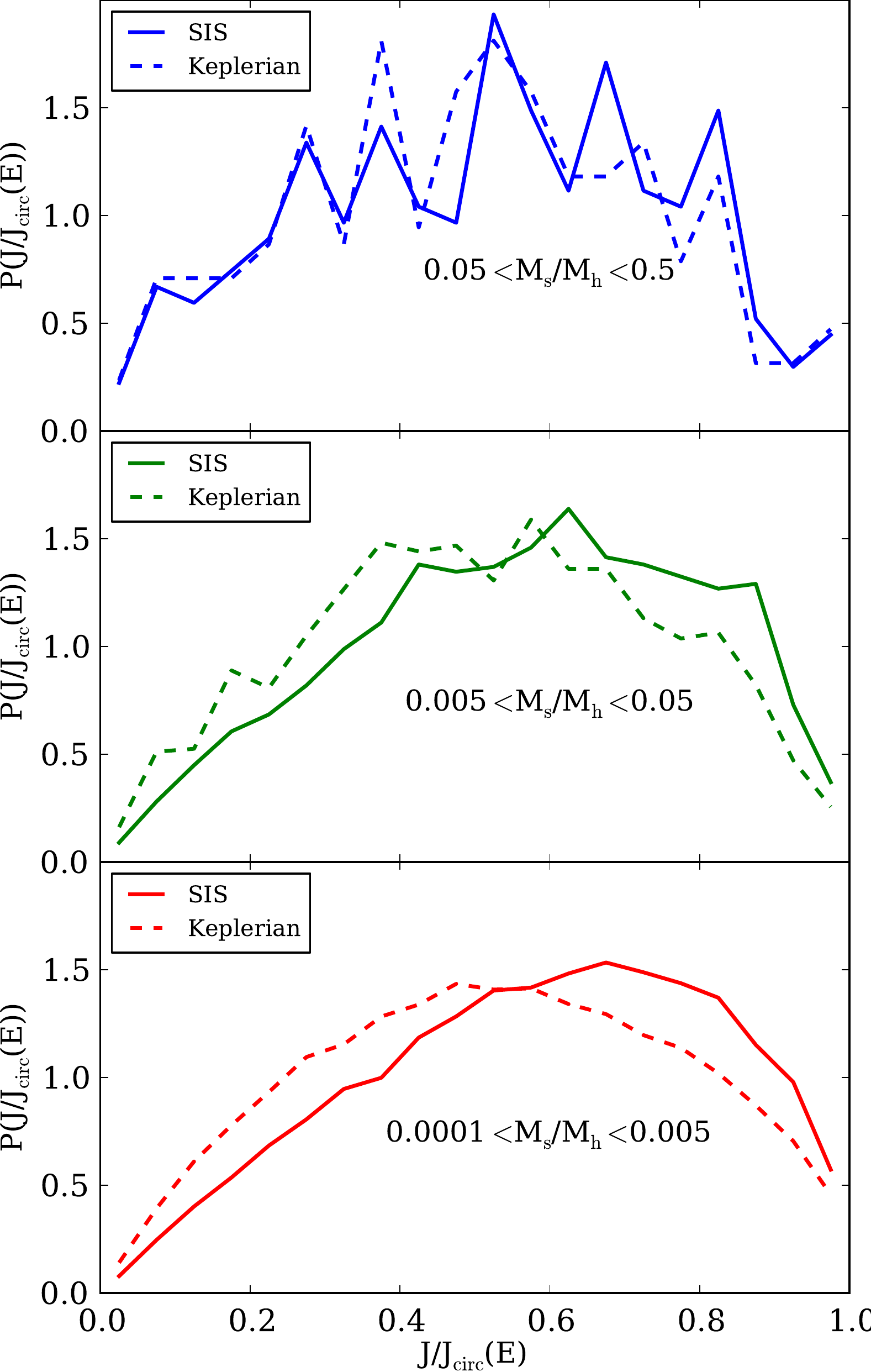}\\%
\caption{Distributions of circularity, $J/J_{\rm circ}(E)$, for
  infalling satellite haloes for host haloes in a mass bin centred on
  $10^{13}$ M$_\odot$.  Solid curves show the distribution derived
  assuming a singular isothermal sphere and dashed curves show the
  distribution derived using the Keplerian model. The three panels are
  for the same three bins of $M_{\rm s}/M_{\rm h}$ as in
  Fig.~\ref{fig:A1}. }
\label{fig:A2}
\end{figure}

The solid curves in Fig.~\ref{fig:A1} compare, as a function of
satellite infall velocity, $V$, the SIS expression with the Keplerian
expression evaluated in the limit $M_{\rm s}/M_{\rm h} \ll 1$, such
that $\mu \rightarrow M_{\rm s}$. The individual points on the
different panels show the results of the full Keplerian expression
with its dependence on $M_{\rm s}/\mu$ applied to our satellite sample
in different bins of $M_{\rm s}/M_{\rm h}$.  The model curves
necessarily agree at $V=V_{200}$ because the mass enclosed in a
circular orbit at $r_{200}$, where the circular velocity is $V_{200}$,
is the same by construction.  For $M_{\rm s}/M_{\rm h} \ll 1$, the
difference between the two models is largest at large $V/V_{200}$
where the orbits extend far beyond $r_{200}$ and hence the mass
enclosed in the SIS greatly exceeds the mass assumed in the point mass
approximation. The effect of the reduced mass, $\mu$, is small for
$M_{\rm s}/M_{\rm h} < 0.05$, but for $0.05<M_{\rm s}/M_{\rm h}<0.5$
it has the effect of reducing $J^{\rm Kep}_{\rm circ}(E)$ and produces
values closer to the SIS case. This is demonstrated in
Fig.~\ref{fig:A2} which compares the distribution of circularities,
$J/J_{\rm circ}(E)$, evaluated using the two different expressions for
three ranges in satellite-to-host mass ratio.  Overall, the two models
agree well with each other for higher values of $M_{\rm s}/M_{\rm h}$,
but they differ for the lowest mass ratio bins.

\label{lastpage}
\end{document}